\documentclass[12pt]{article} 
\usepackage{epsfig,amsmath,subfigure}
\renewenvironment{itemize}
  {\begin{list}%
     {}%
     {\setlength{\topsep}{-6pt}%
      \setlength{\partopsep}{-6pt}%
      \setlength{\itemsep}{-3pt}%
      \setlength{\labelsep}{5pt}%
      \setlength{\itemindent}{0pt}%
     }%
  }%
  {\end{list}}%

\def\cD{{\cal D}}

\def\ren{{\cal R}}

\def\A{{\rm A}}

\def\nn{{\rm n}}

\def\rP{{\rm P}}
\def\T{{\rm T}}
\def\traza{{\rm Tr}}
\def\U{{\rm U}}

\def\ee{\varepsilon}

\def\aslash{{a\mkern-9mu/}}
\def\baslash{\bar{a\mkern-9mu/}}
\def\Aslash{{A\mkern-11mu/}}

\def\Dirac{{D\mkern-12mu/}}
\def\bDirac{\bar{D\mkern-11mu/}}
\def\prslash{{\partial\mkern-9mu/}}
\def\poslash{{p\mkern-8mu/}_1{\!}}

\def\prslash{{\partial\mkern-9mu/}}    
\def\qslash{{q\mkern-8mu/}{\!}}
\def\pislash{{p\mkern-8mu/}_i{\!}}

\def\pone{p_1}
\def\ptwo{p_2}
\def\pthree{p_3}
\def\muo{\mu_1}
\def\mutw{\mu_2}
\def\muth{\mu_3}
\def\mufo{\mu_4}

\def\pt{\tilde p}

\def\idp{\int\! \frac{d^4\!p}{(2\pi)^4} \,\,}

\def\idq{\int\! \frac{d^4\!q}{(2\pi)^4} \,\,}

\def\idpo{\int\! \frac{d^{2\omega}\!p}{(2\pi)^{2\omega}} \,\,}

\def\idqo{\int\! \frac{d^{2\omega}\!q}{(2\pi)^{2\omega}} \,\,}
\def\idx{\int\! d^4\!x \,}

\def\idxo{\int\! d^{2\omega}\!x \,}

\begin{document}
\begin{titlepage}
\rightline{UCM-FT/00-67-2002}

\vskip 1.5 true cm
\begin{center}
{\Large \bf The Gauge Anomaly and the Seiberg-Witten Map}\\ 
\vskip 1.2 true cm 
{\rm C.P. Mart\'{\i}n}\footnote{E-mail: carmelo@elbereth.fis.ucm.es}
\vskip 0.3 true cm
{\it Departamento de F\'{\i}sica Te\'orica I}\\
{\it Facultad de Ciencias F\'{\i}sicas}\\ 
{\it Universidad Complutense de Madrid}\\
{\it 28040 Madrid, Spain}\\
\vskip 1.2 true cm

{\leftskip=45pt \rightskip=45pt 
\noindent
The consistent form of the gauge anomaly is worked out at first order
in $\theta$ for the noncommutative three-point function of the 
ordinary gauge field of  certain noncommutative chiral gauge 
theories defined by  means of the Seiberg-Witten map. We obtain 
that for any compact simple Lie group the anomaly cancellation 
condition of this three-point function reads  
$\traza\, \T^a\,\T^b\,\T^c = 0$, if one restricts the type of noncommutative 
counterterms that can be added to the classical 
action to restore the gauge symmetry to those which are renormalizable by 
power-counting. On the other hand, if the power-counting remormalizability 
paradigm is relinquished and one admits noncommutative counterterms (of the 
gauge fields, its derivatives and $\theta$) which are not power-counting 
renormalizable, then, the anomaly  
cancellation condition for the noncommutative three-point function of the 
ordinary gauge field becomes the ordinary  
one: $\traza\, \T^a\,\{\T^b,\T^c\} = 0$.  
\par }
\end{center}

\vfil

\end{titlepage}
\setcounter{page}{2}


\section{Introduction}

The Seiberg-Witten map was introduced in ref.~\cite{Seiberg:1999vs} to 
account, at least formally, for the physical equivalence of two formulations 
of the same theory. The authors of ref.~\cite{Seiberg:1999vs} studied how 
noncommutative gauge fields and ordinary gauge fields arise
in open string theory for $\U(\nn)$ groups. They showed that either type 
of gauge field can be obtained from the same world-sheet field theory by 
changing the regularization prescription. Since Physics cannot depend on the 
choice of regularization, and a change of the renormalization conditions on 
the string world-sheet corresponds to a field redefinition in space-time, 
Seiberg and Witten concluded that, generally, there must exist a map from 
the ordinary gauge field to its noncommutative counterpart intertwining 
with the gauges symmetries. However, this map does fail to exist in some 
instances~\cite{Seiberg:1999vs, Bytsko:2000di}. Then, the authors of 
refs.~\cite{Madore:2000en, Jurco:2000ja, Jurco:2001my, Jurco:2001rq} 
realized that one can take further advantage of the idea embodied in 
Seiberg-Witten map that a noncommutative gauge field can be defined in terms of its ordinary counterpart, and  
formulated gauge theories on noncommutative space-time for groups other than 
$\U(\nn)$; actually, for arbitrary gauge groups. Thus the 
standard model and GUTs were formulated at the tree level on 
noncommutative space-time~\cite{Calmet:2001na, Aschieri:2002mc}. After a 
promising start~\cite{Bichl:2001cq}  it turned out   
that the noncommutative gauge theories so defined might not be power-counting 
renormalizable in perturbation theory~\cite{Wulkenhaar:2001sq}. And yet, they  
may be phenomenologically useful if, as suggested in ref.~\cite{Calmet:2001na}, 
one embraces the effective field theory philosophy --see refs.~\cite{Dobado:jx, Pich:1998xt} for introductions to effective  field theory. Or, it may well be that supersymmetry~\cite{Bichl:2002wb, Putz:2002ib}  turned these models into 
power-counting renormalizable models in the perturbative expansion.

Several issues concerning the Seiberg-Witten map and the noncommutative field 
theories obtained by using it have been studied in the literature so far. The
perturbative --in $\theta$-- solution to the differential equation defining the
Seiberg-Witten map has been obtained  by  employing several methods in  
refs.~\cite{Okuyama:1999ig, Jurco:2001rq, Garousi:1999ch, Mehen:2000vs, 
Fidanza:2001qm}.
An exact expression for the inverse of Seiberg-Witten map was conjectured in  
ref.~\cite{Liu:2000mj}. It was shown    
in refs.~\cite{Okawa:2001mv, Mukhi:2001vx, Liu:2001pk} that the conjecture is 
correct. In refs.~\cite{Brace:2001fj, Picariello:2001mu, Barnich:2002pb}  
cohomological approaches to the Seiberg-Witten map were put forward. These approaches can be
used to discuss the ambiguities affecting the Seiberg-Witten map which were
pointed out in ref.~\cite{Asakawa:1999cu}. It turns out that the Seiberg-Witten
map is unique modulo gauge transformations and field redefinitions. This
arbitrariness in the value of the Seiberg-Witten map is of the utmost 
importance  in the the renormalization process~\cite{Bichl:2001cq}. 
As happens with the ordinary gauge anomaly, the Seiberg-Witten map also 
involves a (noncommutative) gauge group  
cocycle~\cite{Jurco:2001kp,Jackiw:2002au}. How the Seiberg-Witten map acts
on topological nontrivial noncommutative gauge field configurations has been 
studied by several  
authors~\cite{Hashimoto:2001pc, Kraus:2001xt, Polychronakos:2002pm}. It 
so happens that noncommutative configurations constructed using projection 
operators map to ``commutative'' configurations that have delta-function 
singularities. Thus  it can be exhibited that the physics of noncommutative 
gauge theories is rather different from that of their ordinary counterparts. On the  
phenomenological side the Seiberg-Witten map has been used to generate theories
which lack, due the noncommutativity of space-time, particle Lorentz  
invariance~\cite{Colladay:1998fq}.  Computations of the strength of the
breaking of particle Lorentz invariance has led upon comparison with 
experimental data to  bounds on the scale of the noncommutative 
parameter~\cite{Carroll:2001ws, Carlson:2001sw}. However, no study of the 
gauge anomaly problem have we found in the literature   
--see~\cite{Banerjee:2001un}  for the axial anomaly-- in spite of 
its implications for model building as well as its bearing on the 
quantum consistency of chiral gauge theories in general. We shall try 
to remedy this situation in this paper.

The purpose of this article is to analyze the behaviour under gauge 
transformations of noncommutative gauge theories with chiral fermions   
carrying arbitrary finite dimensional unitary representations of compact 
simple Lie groups. Hence, the formalism put forward in 
 refs.~\cite{Madore:2000en, Jurco:2000ja, Jurco:2001my, Jurco:2001rq} must be
employed and use the Seiberg-Witten map to express the noncommutative fields
in terms of their ordinary counterparts. We shall consider a 
noncommutative left-handed spinor whose ordinary counterpart carries an
arbitrary finite dimensional unitary representation  a compact simple Lie 
group, the generalization to more general instances being straightforward. We 
shall quantize the spinor field and keep the gauge field as a background  
field. 

The lay out of this paper is as follows.  In the first section 
we formulate our model and define a regularized action in terms of the 
ordinary fields. This action is obtained by applying the Seiberg-Witten map  
to an action written in terms of noncommutative fields. Thus the relation 
between noncommutative  and ordinary fields established by the Seiberg-Witten 
map will not be spoiled by the regularization process. Section two is 
devoted to the diagrammatic computation of the anomaly carried by the 
noncommutative three-point function of the ordinary gauge field. This anomaly 
is the noncommutative sibling of the ordinary triangle gauge anomaly. 
In section three, we show by using a mixture of path integral and diagrammatic 
arguments what the consistent form of the gauge anomaly is at first 
order in $\theta$. We shall close the paper with a section in which comments 
and conclusions shall be  given and an Appendix.

\section{The model and its regularization}

Let $G$ be a compact simple Lie group. Let $\psi_L$ denote a left-handed 
spinor on Minkowski space carrying a given finite dimensional 
unitary representation of $G$. Let $a_{\mu}$ denote the gauge field which couples 
to $\psi_L$.  Then, the action that gives the interaction between 
$\psi$ and  $a_{\mu}$ reads
\begin{equation}
S\,=\,\idx\,\bar{\psi}_L\, i\Dirac\,(a)\psi_{L}.
\label{classact}
\end{equation}
The symbol $i \Dirac\,(a)=i\gamma^{\mu}\,D_{\mu}(a)$ denotes the Dirac operator, with $D_{\mu}(a)$ being the covariant derivative:
$D_{\mu}(a)\psi_L =\partial_{\mu}\psi_L-i a_{\mu}\psi_L$. The gamma matrices, 
$\gamma^{\mu}$, $\mu=0,1,2,3$ are defined by 
$\{\gamma^{\mu},\gamma^{\nu}\}=\eta^{\mu\nu}$; $\eta_{\mu\nu}$ being 
the Minkowski metric with $\eta_{00}=1$. As usual, $\bar{\psi}_L=\psi_L^{\dagger}\gamma^{0}$.

The action in eq.~(\ref{classact}) is invariant under the following BRS 
transformations
\begin{equation}
s a_{\mu}=\cD_{\mu}(a)\lambda,\quad s \psi_L=i\lambda\psi_L\quad\mbox{and}\quad s\lambda=i\lambda\lambda.
\label{BRStrans}
\end{equation}
$s$ is the BRS operator, which is linear, commutes with $\partial_{\mu}$, satisfies the anti-Leibniz rule and is  nilpotent ($s^2=0$). 
 $\cD_{\mu}(a)$ is equal to $\partial_{\mu}\,-\,i[a_{\mu},\;]$ and
$\lambda$ denotes the ghost field, which has ghost number 1. Both $a_{\mu}$ and $\psi$ have ghost number 0.
 
To construct the noncommutative counterpart of the ordinary theory defined by 
$S$, we shall employ the formalism developed in refs.~\cite{Madore:2000en, Jurco:2000ja, Jurco:2001my, Jurco:2001rq}. Let $A_{\mu}$, $\Psi_L$ and $\Lambda$ 
stand for the noncommutative gauge field, the noncommutative left-handed 
spinor field and the noncommutative ghost field, respectively.  
$A_{\mu}$, $\Psi_L$ and $\Lambda$ are defined 
in terms of $a_{\mu}$, $\psi_L$ and $\lambda$ by means of the Seiberg-Witten
map. Modulo BRS transformations and field redefinitions, the Seiberg-Witten
map at first order in $\theta$ reads
\begin{equation}
\begin{array}{l}
{A_{\mu}(a,\theta)\,=\,a_\mu\,-\,\frac{1}{4}\,
\theta^{\alpha\beta}\,
\{a_{\alpha},f_{\beta\mu}\,+\,\partial_{\beta} a_{\mu}\}\,+\,o(\theta^2),}\\[9pt]
{\Psi_L(a,\psi_L,\theta)\,=\,\psi_L-\frac{1}{2}\,
\theta^{\alpha\beta}\,a_{\alpha}\partial_{\beta}\psi_L\,+
\,\frac{i}{8}\,\theta^{\alpha\beta}\,[a_{\alpha},a_{\beta}]\,\psi_L
\,+\,o(\theta^2),}\\[9pt]
{\Lambda(a,\lambda,\theta)\,=\,\lambda\,+\,\frac{1}{4}\,\theta^{\alpha\beta}\,\{\partial_{\alpha}\lambda,a_{\beta}\}\,+\,o(\theta^2).}\\[9pt]
\end{array}
\label{SeibergWitten}
\end{equation}
Note that $A_{\mu}$ and $\Lambda$ are valued in the representation of the
enveloping algebra of the Lie algebra of $G$ induced by the unitary  
representation of the latter algebra carried by $\psi_L$~\cite{Jurco:2000ja}.

Let $\star$ denote the Moyal product of  functions on Minkowski space:
\begin{displaymath}
(f\star g)(x)\,=\,\idp\,\idq e^{-i(p+q)x}\,
e^{-\frac{i}{2}\theta^{\alpha\beta}p_{\alpha}q_{\beta}}\,\tilde{f}(p)
\,\tilde{q}(q);
\end{displaymath}
$\tilde{f}(p)$ and $\tilde{q}(q)$ being the Fourier transforms of $f$ and $g$,
respectively. Then, the noncommutative, $S_{nc}$, version of the action in 
eq.~(\ref{classact}) reads
\begin{equation}
S_{nc}\,=\,\idx\,\bar{\Psi}_L\star i\Dirac\,(A)\Psi_{L},
\label{ncclassact}
\end{equation}
where  $\Dirac\,(A)=\gamma^{\mu}D_{\mu}(A)$; $D_{\mu}(A)$ being the 
noncommutative covariant derivative:  
$D_{\mu}(A)\Psi_L =\partial_{\mu}\Psi_L-i A_{\mu}\star\Psi_L$. Again,  
$\bar{\Psi}_L=\Psi_L^{\dagger}\gamma^{0}$.

The noncommutative action $S_{nc}$ is invariant under the action of 
the noncommutative BRS operator $s_{\star}$. $s_{\star}$ is a linear 
operator which commutes with $\partial_{\mu}$, satisfies the anti-Leibniz 
rule and acts on the noncommutative fields as follows 
\begin{equation}
s_{\star}\,A_{\mu}=\cD_{\mu}(A)\,\Lambda,\quad s_{\star} \Psi_L=i\Lambda \star\Psi_L\quad\mbox{and}\quad s_{\star} \Lambda=i\Lambda\star\Lambda.
\label{NCBRStrans}
\end{equation}
The symbol $\cD_{\mu}(A)$ stands for $\partial_{\mu}-i[A_{\mu},\;]_{*}$;
$[f,g]_{*}=f\star g -g\star f$. $s_{\star}$ is  nilpotent.

By definition of  Seiberg-Witten map, the following equations 
hold~\cite{Brace:2001fj}
\begin{equation}
s A_{\mu}(a,\theta)\,=\,s_{\star} A_{\mu} \quad 
s \Psi_L(a,\psi_L,\theta)\,=\,s_{\star} \Psi_L\quad\mbox{and}\quad 
s \Lambda(a,\lambda,\theta)\,=\,s_{\star} \Lambda.
\label{SWequations}
\end{equation}
The action of $s$ on $A_{\mu}(a,\theta)$, $\Psi_L(a,\psi_L,\theta)$ and  
$\Lambda(a,\lambda,\theta)$ is computed by assuming first that these objects 
are formal power series of $\theta$, the ordinary fields and 
their derivatives, and, then, using eq.~(\ref{BRStrans}). The right hand side 
of the identities in eq.~(\ref{SWequations}) is given by eq.~(\ref{NCBRStrans}).

By expanding $S_{nc}$ in terms of $a$ and $\psi_L$ to first order in $\theta$,
one obtains 
\begin{equation}
S_{nc}=\idx\,\bar{\psi}_L \{i\Dirac\,(a) -\frac{1}{2}\theta^{\alpha\beta}
[\frac{1}{2}f_{\alpha\beta} i\Dirac\,(a)+\gamma^{\rho}
f_{\rho\alpha}iD_{\beta}(a)]\}
\psi_L\,+\, o(\theta^2).
\label{expanone}
\end{equation}
From now on, we shall use the notation 
$f_{\alpha\beta}=\partial_{\alpha}a_{\beta}-\partial_{\beta}a_{\alpha}
-i\,[a_{\alpha},a_{\beta}]$.

Upon quantizing $\psi_L$, the previous  action  can be used to define a 
noncommutative quantum field theory on a background gauge field.
For technical reasons --we will employ dimensional regularization as defined
in ref.~\cite{Breitenlohner:hr} and thus we shall need the Dirac propagator 
for describing the free propagation of the fermionic degrees of 
freedom--, we shall use an action which gives the very same interacting 
theory between gauge  and  
fermion fields as  the action of eq.~(\ref{ncclassact}), but whose 
kinetic term is that of a Dirac spinor. Let $\psi$  denote an ordinary Dirac 
spinor such that $\psi_L\,=\,\frac{1}{2}(1-\gamma_5)\psi$, where $\psi_L$ is 
the left-handed spinor introduced at the beginning of this section
and $\gamma_5=i\gamma^0\gamma^1\gamma^2\gamma^3$. We shall 
define the action on $\psi$ of the BRS operator $s$ as follows
\begin{equation}
s\psi\,=\,i\lambda{\rm P}_{-}\psi,
\label{BRSchiral}
\end{equation}
where ${\rm P}_{-}=\frac{1}{2}(1-\gamma_5)$.

Let $\Psi$ be a noncommutative spinor which is a solution to the following 
Seiberg-Witten problem:
\begin{equation}
\begin{array}{l}
{s \Psi(a,\psi,\theta)\,=\,i\Lambda\star {\rm P}_{-}\Psi,}\\[9pt]
{\Psi(a,\psi,\theta=0)=\psi.}\\[9pt]
\end{array}
\label{Diracseibergeq}     
\end{equation}
Modulo BRS transformations and field redefinitions, the solution to the previous equation reads
\begin{equation}
\Psi(a,\psi,\theta)\,=\,\psi-\frac{1}{2}\,
\theta^{\alpha\beta}\,a_{\alpha}{\rm P}_{-}\partial_{\beta}\psi\,+
\,\frac{i}{8}\,\theta^{\alpha\beta}\,[a_{\alpha},a_{\beta}]\,{\rm P}_{-}\psi
\,+\,o(\theta^2).
\label{swmappsi}
\end{equation}

Let the noncommutative action describing the interaction between $\Psi$ 
in eq.~(\ref{swmappsi}) and $A_{\mu}$ in eq.~(\ref{SeibergWitten}) be given
by 
\begin{equation}
S^{(-)}_{nc}\,=\,\idx\,\bar{\Psi}\star i\hat{D}(A)\Psi.
\label{ncaction}
\end{equation}
The symbol $\hat{D}(A)$ denotes the following operator 
$\hat{D}(A)\Psi\,=\,\prslash\Psi-i\Aslash\star\,{\rm P}_{-}\Psi$. 

The action in eq.~(\ref{ncaction}), with $A_{\mu}$ and $\Psi$ given in
eqs.~(\ref{SeibergWitten}) and~(\ref{swmappsi}), defines the same interacting 
theory as  the action in eq.~(\ref{classact}), 
with $A_{\mu}$ and $\Psi_L$ as in  eq.~(\ref{SeibergWitten}), since $\psi_R=\frac{1}{2}(1+\gamma_5)\psi$ 
does not couple to the gauge field $a_{\mu}$. Up to first order in $\theta$, 
$S^{(-)}_{nc}$ reads
\begin{equation}
S^{(-)}_{nc}=\idx\,\bar{\psi} \{i\prslash\psi 
+\aslash\,{\rm P}_{-}-\frac{1}{2}\theta^{\alpha\beta}
[\frac{1}{2}f_{\alpha\beta} i\Dirac\,(a)\,{\rm P}_{-}+
\gamma^{\rho}f_{\rho\alpha}iD_{\beta}\,(a)\,{\rm P}_{-}]\}
\psi\,+\, o(\theta^2).
\label{expantwo}
\end{equation}
Note that if we do perturbation theory with $S^{(-)}_{nc}$, the free propagator 
for $\psi$ is that of Dirac's. Also note that, as in ordinary Quantum Field
Theory~\cite{Alvarez-Gaume:1983cs}, one can use $S^{(-)}_{nc}$ to define the 
Wick rotated counterpart of the path integral
\begin{displaymath}
\int d\bar{\psi}\,d\psi\quad e^{\,iS^{(-)}_{nc}}
\end{displaymath} 
as the determinant of the operator 
\begin{displaymath}
{\cal O}=i\prslash+\aslash\,{\rm P}_{-}-\frac{1}{2}\theta^{\alpha\beta}
[\frac{1}{2}f_{\alpha\beta} i\Dirac\,(a)\,{\rm P}_{-}+
\gamma^{\rho}f_{\rho\alpha}iD_{\beta}\,(a)\,{\rm P}_{-}].
\end{displaymath}
${\cal O}$ has a well-defined eigenvalue problem, at least at first 
order in $\theta$, over Dirac spinors on Euclidean space. Let us remark that 
if we had used $S_{nc}$ in eq.~(\ref{expanone}) instead of $S^{(-)}_{nc}$,
this definition of the path integral of the theory would have had no meaning:
the operator in $S_{nc}$ maps left-handed spinors to right-handed spinors 
so that its Euclidean version has no eigenvalue problem. 

Now we come to one of chief issues in this paper: the choice of a 
regularization that does not spoil the noncommutative origin of the theory
whose action is $S^{(-)}_{nc}$ in eq.~(\ref{expantwo}). Since we lack a
characterization of the noncommutative origin of the theory that only involved
the ordinary fields $a_{\mu}$ and $\psi$ --e.g., some equation to be 
satisfied by the 1PI functional of the theory when expressed in terms
of the ordinary fields--, the best we can do is to formulate an
action in terms of the noncommutative  fields which yields upon  
application of the Seiberg-Witten map a regularized 
action --i.e., an action which gives rise to regularized Feynman diagrams 
of $a_{\mu}$ and $\psi$. We shall do this by using Dimensional Regularization 
as systematized by the authors of ref.~\cite{Breitenlohner:hr}. We shall thus 
use a non-anticommuting $\gamma_{5}$ object and employ the ``hat-and-bar'' 
notation of ref.~\cite{Breitenlohner:hr} --see also  
refs.~\cite{Martin:1999cc, Sanchez-Ruiz:2002xc}.
 
We shall define the object $\theta^{\mu\nu}$  in dimensional regularization 
as an algebraic object which satisfies
\begin{displaymath}
\theta^{\mu\nu}=-\theta^{\nu\mu},\,\quad \theta^{\mu\rho}\hat{g}_{\rho\nu}=0.
\end{displaymath}
We introduce now the noncommutative regularized action
\begin{equation} 
S^{(-)}_{nc,\,DR}\,=\,
\idxo\,
\bar{\Psi}\star\{\prslash\Psi-iA_{\mu}\bar{\gamma}^{\mu}\star\,P_{-}\Psi\}.
\label{ncDR}
\end{equation}
Let us generalize next to the $2\omega$-dimensional space of Dimensional 
Regularization the BRS transformations in four dimensions of 
$a_{\mu}$, $\lambda$, $\psi$, $A_{\mu}$, $\Lambda$ and $\Psi$ 
--see eqs.~(\ref{BRStrans}), (\ref{NCBRStrans}), (\ref{BRSchiral}) 
and~(\ref{Diracseibergeq}). We shall choose a straightforward 
generalization of the latter so that the BRS transformations  look the 
same in ``$2\omega$-dimensions''  as  in four. Hence, the Seiberg-Witten 
equations in the $2\omega$-dimensional space of dimensional regularization   
read 
\begin{equation}
s A_{\mu}\,=\,\cD_{\mu}(A)\Lambda,  \quad
s \Psi\,=\,i\Lambda\star {\rm P}_{-}\Psi\quad\mbox{and}\quad 
s \Lambda\,=\,i\Lambda\star \Lambda,
\label{DRSWeq}
\end{equation}
where $A_{\mu}=A_{\mu}(a,\theta)$, $\Psi=\Psi(a,\psi,\theta)$ and 
$\Lambda=\Lambda(a,\lambda,\theta)$, and $s$ acts on the ordinary fields in
``$2\omega$-dimensions'' as it does on their counterparts in four dimensions:
\begin{equation}
s a_{\mu}\,=\,\cD_{\mu}(a)\lambda,\,  \quad
s \psi\,=\,i\lambda{\rm P}_{-}\psi\quad\mbox{and}\quad 
s \lambda\,=\,i\lambda \lambda.
\label{DRBRSeq}
\end{equation}
The previous Seiberg-Witten equations --eq.~(\ref{DRSWeq})-- solved 
for the appropriate boundary  conditions 
--i.e., $A_{\mu}(a,\theta=0)=a_{\mu}$,  
$\Psi(a,\psi,\theta=0)=\psi$ and  $\Lambda(a,\lambda,\theta=0)=\lambda$-- 
yield the Seiberg-Witten map in the $2\omega$-dimensional space of 
Dimensional Regularization. It is apparent that modulo field redefinitions and
BRS transformations the Seiberg-Witten map obtained from eqs.~(\ref{DRSWeq}) 
will look the same as in four dimensions:
\begin{equation}
\begin{array}{l}
{A_{\mu}(a,\theta)\,=\,a_\mu\,-\,\frac{1}{4}\,
\theta^{\alpha\beta}\,
\{a_{\alpha},f_{\beta\mu}\,+\,\partial_{\beta} a_{\mu}\}\,+\,o(\theta^2),}\\[9pt]
{\Psi(a,\psi,\theta)\,=\,\psi -\frac{1}{2}\,
\theta^{\alpha\beta}\,a_{\alpha}{\rm P}_{-}\partial_{\beta}\psi\,+
\,\frac{i}{8}\,\theta^{\alpha\beta}\,[a_{\alpha},a_{\beta}]\,{\rm P}_{-}\psi
\,+\,o(\theta^2),}\\[9pt]
{\Lambda(a,\lambda,\theta)\,=\,\lambda\,+\,\frac{1}{4}\,\theta^{\alpha\beta}\,\{\partial_{\alpha}\lambda,a_{\beta}\}\,+\,o(\theta^2).}\\[9pt]
\end{array}
\label{DRSWmap}
\end{equation}
Every  object in the previous equations is an  algebraic object in the 
$2\omega$-dimensional space of dimensional regularization. 

Now, substituting eq.~(\ref{DRSWmap}) in eq.~(\ref{ncDR}) and expanding 
at first order in $\theta$, one obtains a regularized version of
$S^{(-)}_{nc}$ in eq.~(\ref{expantwo}): 
\begin{equation}
\begin{array}{l}
{S^{(-)}_{nc,\,DR}=\bar{S}_{nc}\,+\,\hat{S}_{nc},}\\[9pt]
{\bar{S}_{nc}=\idxo\,\bar{\psi} \{i\prslash\psi 
+\baslash\,{\rm P}_{-}-\frac{1}{2}\theta^{\alpha\beta}
[\frac{1}{2}f_{\alpha\beta} i \bDirac\,(a)\,{\rm P}_{-}+
\bar{\gamma}^{\rho}f_{\rho\alpha}iD_{\beta}(a)\,{\rm P}_{-}]\}
\psi}\\[9pt]
{\hat{S}_{nc}=-\frac{i}{2}\theta^{\alpha\beta}
\idxo\,\bar{\psi}[\partial_{\alpha}a_{\beta}+ a_{\beta}\partial_{\alpha}  
-\frac{i}{2}a_{\alpha}a_{\beta}]\,\hat{\prslash}\,{\rm P}_{+}\psi ,}\\[9pt]
{\phantom{\hat{S}_{nc}=}
+\frac{i}{2}\theta^{\alpha\beta}
\idxo\,\bar{\psi}[\hat{\prslash}a_{\beta}\partial_{\alpha}
+\frac{i}{2}(\hat{\prslash}a_{\alpha}a_{\beta}+
                   a_{\alpha}\hat{\prslash}a_{\beta})+
(a_{\beta}\partial_{\alpha}+\frac{i}{2}
a_{\alpha}a_{\beta})\hat{\prslash}\,]\,{\rm P}_{-}\psi.}\\[9pt]
\end{array}
\label{DRexpanded}
\end{equation}
We have used the following notation: $\baslash=a_{\mu}\bar{\gamma}^{\mu}$,
$\bDirac\,(a)=\bar{\gamma}^{\mu}D_{\mu}(a)$ and
$\hat{\prslash}=\hat{\gamma}^{\mu}\partial_{\mu}$.

Furnished with the action in eq.~(\ref{DRexpanded}) and  employing standard 
Feynman diagram techniques, we can set up a dimensionally regularized perturbative quantum field theory. Explicit computations will be carried out  below. 

Before we close this section we would like to make two comments. 
First,  let us  stress that by using a regularized action for the 
ordinary fields --the action in eq.~(\ref{DRexpanded})-- which comes via the 
Seiberg-Witten map from a noncommutative 
object --the action in eq.~(\ref{ncDR})--, we make
sure that the regularization method does not erase (partially or totally) the 
noncommutative origin of our theory. If the regularized action in terms of 
the ordinary fields could not  be obtained via the Seiberg-Witten map
from a noncommutative object, then there would be no guarantee that the 
renormalized theory  based on that regularized action would have a 
noncommutative interpretation. However, it might well happen --as it the case in the algebraic renormalization of gauge theories-- that by adding 
appropriate counterterms to the action of this 
renormalized theory, a  theory having a noncommutative content can 
be worked out. Probably, the counterterms needed to restore, and compensate for
the lack of it,  the  noncommutative origin of the theory 
will not have a noncommutative content. Of course, anomalies in 
the Seiberg-Witten map might arise, if no regularized action with a 
noncommutative interpretation is found. 
To settle all these issues in a regularization independent way, we need an 
equation (or equations) involving only the ordinary fields  which tells us 
when a given 1PI functional  defines a theory having a noncommutative origin. 
We lack such an equation (or set of equations); so, for the time being the
only way to proceed is as it is done in this paper. Second, the dimensionally
regularized theory defined by the action in eq.~(\ref{DRexpanded}) is not
BRS invariant since $s$ in eq.~(\ref{DRBRSeq}) fails to annihilate 
$S^{(-)}_{nc,\,DR}$. This befits the occurrence of gauge anomalies. We shall 
begin to explore the consequences that  
this BRS-breaking brings about in the next section.

\section{The anomaly in the three-point function of the gauge field}

Let $\Gamma[a,\theta]$ be the renormalized noncommutative effective action 
for the ordinary gauge field which is defined as follows
\begin{displaymath}
e^{\,i \Gamma[a,\theta]}\,=\,\ren\,
\Big\{\int\,d\bar{\psi}\,d{\psi}\;e^{\,iS^{(-)}_{nc,\,DR}}\Big\}.
\end{displaymath}
$\ren\{\cdots\}$ stands for renormalization of the sum of
dimensionally regularized diagrams represented by the formal path integral 
between the curly brackets.
 
Since $S^{(-)}_{nc}$ in eq.~(\ref{expantwo}) is BRS invariant, one aims 
at constructing --by choosing appropriate counterterms-- a 
$\Gamma[a,\theta]$ which be BRS invariant as well, i.e.,
\begin{equation}
s\Gamma[a,\theta]\,=\,0.
\label{BRSGamma}
\end{equation}
See eqs.~(\ref{BRStrans}) and ~(\ref{BRSchiral}) for the definition of $s$.
We shall show below that eq.~(\ref{BRSGamma}) cannot hold 
for any finite dimensional unitary representation of any compact simple 
Lie group, if only noncommutative  power-counting renormalizable 
counterterms are allowed. But we
shall also show that if power-counting renormalizability is given up and 
the ordinary anomaly cancellation condition~\cite{Gross:pv} is satisfied, 
then, at least the noncommutative three-point function of $a_{\mu}$ is 
anomaly free.

For the Fourier transform of  the two-point and three-point functions,
eq.~(\ref{BRSGamma}) boils down to
\begin{equation}
\begin{array}{l}
{a)\quad\quad ip^{\mu}\,\Gamma^{ab}_{\mu\nu}(p)\,=\,0,}\\[9pt]
{b)\quad\quad  i\pthree^{\muth}\,
\Gamma^{a_1 a_2 a_3}_{\muo\mutw\muth}(\pone,\ptwo,\pthree)\,=\,
f^{a_2 a_3 c}\,\Gamma^{a_1 c}_{\muo\mutw}(\pone)-f^{a_3 a_1 c}\,
\Gamma^{a_2 c}_{\mutw\muo}(\ptwo),}\\[9pt]
\end{array}
\label{threepoint}
\end{equation}
with $\pone+\ptwo+\pthree=0$.
Since any contribution involving an even number of $\gamma_5$'s to a  
Feynman diagram can be dimensionally regularized as in a  
vector-like theory, it turns out that only the parity violating contributions  
--contributions with an odd number of $\gamma_5$'s-- to a Feynman diagram can
yield truly anomalous contributions. Hence, if $a)$ and $b)$ in 
eq.~(\ref{threepoint}) are violated, it is the contributions to 
$\Gamma^{ab}_{\mu\nu}(p)$ and 
$\Gamma^{a_1 a_2 a_3}_{\muo\mutw\muth}(\pone,\ptwo,\pthree)$ involving
$\ee_{\muo\mutw\muth\mu_{4}}$ which give rise to  such a breaking 
of BRS invariance. 

Let 
\begin{displaymath}
\Gamma^{(odd)\;ab}_{DR\phantom{d}\,\;\mu\nu}(p)\quad\mbox{and}\quad
\Gamma^{(odd)\;a_1 a_2 a_3}_{DR\phantom{d}\,\;\muo\mutw\muth}
(\pone,\ptwo,\pthree)
\end{displaymath}
denote, respectively, the dimensionally regularized  contributions to the 
noncommutative two-point and three-point functions of $a_{\mu}$ which 
depend on   
$\ee_{\muo\mutw\muth\mu_{4}}$. These regularized contributions are
calculated with the action in eq.~(\ref{DRexpanded}). We have found
that up to first order in $\theta$ the two-point function reads
\begin{equation}
\Gamma^{(odd)\;ab}_{DR\phantom{d}\,\;\mu\nu}(p)\,=\,0\,+\,o(\omega-2). 
\label{DRtwopoint}
\end{equation}
As for the three-point function we have obtained the following results
\begin{equation}
\begin{array}{c}
{i\pthree^{\muth}\,
\Gamma^{(odd)\;a_1 a_2 a_3}_{DR\phantom{d}\,\;\muo\mutw\muth}
(\pone,\ptwo,\pthree)=
i\pthree^{\muth}\, 
\Gamma^{(odd)\;a_1 a_2 a_3}_{DR\phantom{d}\,\;\muo\mutw\muth}
(\pone,\ptwo,\pthree)_{triangle}}\\[9pt]
{\phantom{i\pthree^{\muth}\,
\Gamma^{(odd)\;a_1 a_2 a_3}_{DR\phantom{d}\,\;\muo\mutw\muth}(\pone,\ptwo,\pthree)}
+i\pthree^{\muth}\,\Gamma^{(odd)\;a_1 a_2 a_3}_{DR\phantom{d}\,\;\muo\mutw\muth}
(\pone,\ptwo,\pthree)_{swordfish}}\\[9pt]
{\phantom{i\pthree^{\muth}\,
\Gamma^{(odd)\;a_1 a_2 a_3}_{DR\phantom{d}\,\;\muo\mutw\muth}(\pone,\ptwo,\pthree)}
+i\pthree^{\muth}\, 
\Gamma^{(odd)\,a_1 a_2 a_3}_{DR\phantom{d}\,\;\muo\mutw\muth}
(\pone,\ptwo,\pthree)_{jellyfish},}\\[9pt]
\end{array}
\label{DRthreepoint}
\end{equation}
where 
\begin{equation}
\begin{array}{l}
{i\pthree^{\muth}\, 
\Gamma^{(odd)\;a_1 a_2 a_3}_{DR\phantom{d}\,\;\muo\mutw\muth}
(\pone,\ptwo,\pthree)_{triangle}=\traza\Big(\{\T^{a_1},\T^{a_2}\}\T^{a_3}\Big)
\,\frac{i}{24\pi^2}\,
\ee_{\muo\mutw\rho\sigma}\,p_1^{\rho}\,p_2^{\sigma},}\\[9pt]
{
\quad\quad\quad-i\,\traza\Big([\T^{a_1},\T^{a_2}]\T^{a_3}\Big)\Big\{
{\rm I}(p_1^2)\,\big(\ee_{\muo\mutw\rho\sigma}\,p_1^{\rho}\,\pt_3^{\sigma}+
\ee_{\muo\rho\sigma\tau}
\,p_1^{\rho}\,p_2^{\sigma}\,\theta_{\mutw}^{\phantom{\mutw}\,\tau}\big)
}\\[9pt]
{
\phantom{\quad\quad\quad-i\,\traza\Big([\T^{a_1},\T^{a_2}]\T^{a_3}}
+{\rm I}(p_2^2)\,\big(\ee_{\muo\mutw\rho\sigma}\,p_2^{\rho}\,\pt_3^{\sigma}+
\ee_{\mutw\rho\sigma\tau}
\,p_1^{\rho}\,p_2^{\sigma}\,\theta_{\mutw}^{\phantom{\muo}\,\tau}\big)
}\\[9pt]
{\phantom{\quad\quad\quad-i\,\traza\Big([\T^{a_1},\T^{a_2}]\T^{a_3}}
+\frac{i}{48\pi^2}\,
\ee_{\muo\mutw\rho\sigma}\,p_1^{\rho}\,p_2^{\sigma}
\,\theta_{\alpha\beta}\,p_{1}^{\alpha}\,p_{2}^{\beta}\Big\}+o(\omega-2),}\\[9pt]
{
i\pthree^{\muth}\,\Gamma^{(odd)\;a_1 a_2 a_3}_{DR\phantom{d}\,\;
\muo\mutw\muth}(\pone,\ptwo,\pthree)_{swordfish}=
}\\[9pt]
{
\quad\quad\quad+i\,\traza\Big([\T^{a_1},\T^{a_2}]\T^{a_3}\Big)\Big\{
{\rm I}(p_1^2)\,\big(\ee_{\muo\mutw\rho\sigma}\,p_1^{\rho}\,\pt_3^{\sigma}+
\ee_{\muo\rho\sigma\tau}
\,p_1^{\rho}\,p_2^{\sigma}\,\theta_{\mutw}^{\phantom{\mutw}\,\tau}\big)
}\\[9pt]
{
\phantom{\quad\quad\quad-i\,\traza\Big([\T^{a_1},\T^{a_2}]\T^{a_3}}
+
{\rm I}(p_2^2)\,\big(\ee_{\muo\mutw\rho\sigma}\,p_2^{\rho}\,\pt_3^{\sigma}+
\ee_{\mutw\rho\sigma\tau}
\,p_1^{\rho}\,p_2^{\sigma}\,\theta_{\muo}^{\phantom{\muo}\,\tau}\big)\Big\}
+o(\omega-2),}\\[9pt]
{
i\pthree^{\muth}\, 
\Gamma^{(odd)\,a_1 a_2 a_3}_{DR\phantom{d}\,\;\muo\mutw\muth}
(\pone,\ptwo,\pthree)_{jellyfish}=0.}\\[9pt]
\end{array}
\label{results}
\end{equation}
Here, every $o(\theta^2)$-contribution has been dropped and  the following 
shorthand has been adopted  
\begin{equation}
\pt^{\mu}=\theta^{\mu\nu}\,p_{\nu},\quad\quad{\rm I}(p^2)\,=\,\frac{i}{192\pi^2}\Big(\frac{1}{\omega-2}\,+\,
\ln(-\frac{p^2}{4\pi\mu^2})\,+\,\gamma\,-\,\frac{8}{3}\Big)\,p^2.
\label{Ipedos}
\end{equation}
Further details can be found in the Appendix. Eqs.~(\ref{DRthreepoint}) and~(\ref{results}) lead finally to 
\begin{equation} 
\begin{array}{l}
{
i\pthree^{\muth}\,
\Gamma^{(odd)\;a_1 a_2 a_3}_{DR\phantom{d}\,\;\muo\mutw\muth}(\pone,\ptwo,\pthree)=
}\\[9pt]
{

\frac{i}{24\pi^2}\Big(\,\traza\,(\,\{\T^{a_1},\T^{a_2}\}\T^{a_3}\,)\,-\,
\frac{i}{2}\,\theta_{\alpha\beta}\,p_{1}^{\alpha}\,p_{2}^{\beta}\,
\traza\,(\,[\T^{a_1},\T^{a_2}]\T^{a_3}\,)\,\Big)\,\ee_{\muo\mutw\rho\sigma}\,p_1^{\rho}\,p_2^{\sigma}\,+\,O(\omega-2).
}\\[9pt]
\end{array}
\label{triananomaly}
\end{equation}
We should stress that the simplicity of the this eq. belies the complexity of 
the intermediate computations involved in its calculation. It is the fact that 
$S^{(-)}_{nc,\,DR}$ in eq.~(\ref{DRexpanded}) is obtained from 
$S^{(-)}_{nc,\,DR}$ in eq.~(\ref{ncDR})  upon Seiberg-Witten mapping 
--the Seiberg-Witten connection between noncommutative and commutative
fields being thus manifestly preserved-- which should be held responsible 
form the simplicity of the final outcome --eq.~(\ref{triananomaly})-- of our 
computations. Indeed, it can be seen in the Appendix that if we define the
regularized theory by employing just $\bar{S}$ in eq.~(\ref{DRexpanded}) 
instead of using the full $S^{(-)}_{nc,\,DR}$ disaster sets in. The  
$o(\theta)$-contribution to the left hand side of eq.~(\ref{triananomaly}) 
becomes --see Appendix-- the following ugly expression: 
\begin{equation}
\begin{array}{l}
{
\phantom{+\,}\traza\Big(\{\T^{a_1},\T^{a_2}\}\T^{a_3}\Big)
\,\frac{i}{24\pi^2}\,
\ee_{\muo\mutw\rho\sigma}\,p_1^{\rho}\,p_2^{\sigma}
}\\[9pt]
{
+\,\traza\Big([\T^{a_1},\T^{a_2}]\T^{a_3}\Big)
\frac{1}{16\pi^2(\omega-2)}\Big(\frac{1}{24}\Big)
\Big\{
\ee_{\muo\rho\sigma\tau}
\,p_1^{\rho}\,p_2^{\sigma}\,\theta_{\mutw}^{\phantom{\mutw}\tau}\,
(\hat{p}^2_2+\hat{p}^2_3-\hat{p}_2\cdot\hat{p}_3)
}\\[9pt]
{
\phantom{+i\,\traza\Big([\T^{a_1},\T^{a_2}]\T^{a_3}\Big)
\frac{1}{16\pi^2(\omega-2)}\Big(-\frac{i}{24}\,}
+\ee_{\mutw\rho\sigma\tau}
\,p_1^{\rho}\,p_2^{\sigma}\,\theta_{\muo}^{\phantom{\muo}\tau}\,
(\hat{p}^2_1\,+\,\hat{p}^3_2-\hat{p}_1\cdot\hat{p}_3)
}\\[9pt]
{
\phantom{+i\,\traza\Big([\T^{a_1},\T^{a_2}]\T^{a_3}\Big)
\frac{1}{16\pi^2(\omega-2)}\Big(-\frac{i}{24}\,}
+\ee_{\muo\mutw\rho\sigma}\,p_1^{\rho}\,\pt_3^{\sigma}\,
(\hat{p}^2_2+\hat{p}^2_3-\hat{p}_2\cdot\hat{p}_3)
}\\[9pt]
{
\phantom{+i\,\traza\Big([\T^{a_1},\T^{a_2}]\T^{a_3}\Big)
\frac{1}{16\pi^2(\omega-2)}\Big(-\frac{i}{24}\,}
+\ee_{\muo\mutw\rho\sigma}\,p_2^{\rho}\,\pt_3^{\sigma}\,
(\hat{p}^2_1+\hat{p}^2_3-\hat{p}_1\cdot\hat{p}_3)\Big\}
}\\[9pt]
{
-\,\traza\Big([\T^{a_1},\T^{a_2}]\T^{a_3}\Big)
\frac{1}{16\pi^2}\Big(\frac{1}{24}\Big)\Big(
\ee_{\muo\rho\sigma\tau}
\,p_1^{\rho}\,p_2^{\sigma}\,\theta_{\mutw}^{\phantom{\mutw}\,\tau}+
\ee_{\mutw\rho\sigma\tau}
\,p_1^{\rho}\,p_2^{\sigma}\,\theta_{\muo}^{\phantom{\muo}\,\tau}
}\\[9pt]
{
\phantom{-\,\traza\Big([\T^{a_1},\T^{a_2}]\T^{a_3}\Big)
\frac{1}{16\pi^2}\Big(\frac{1}{24}\Big)\Big(
\ee_{\muo\rho\sigma\tau}
\,p_1^{\rho}\,p_2^{\sigma}\,\theta_{\mutw}^{\phantom{\mutw}\,\tau}}
-\ee_{\muo\mutw\rho\sigma}\,p_3^{\rho}\,\pt_3^{\sigma}\Big)
(\bar{p}_1^2+\bar{p}_2^2+\bar{p}_1\cdot\bar{p}_2)
}\\[9pt]
{
+\,\traza\Big([\T^{a_1},\T^{a_2}]\T^{a_3}\Big)
\frac{1}{16\pi^2}\Big(\frac{1}{4}\Big)\,
\ee_{\muo\mutw\rho\sigma}\,p_1^{\rho}\,p_2^{\sigma}
\,\theta_{\alpha\beta}\,p_{1}^{\alpha}\,p_{2}^{\beta}\,+\,o(\omega-2),}\\[9pt]
\end{array}
\label{disaster}
\end{equation}
where $p_1+p_2+p_3=0$, $\hat{p}_i^{\mu}=\hat{g}^{\mu\nu}\,p_{i\,\nu}$,
$i=1,2,3$, $\bar{p}_i^{\mu}=\bar{g}^{\mu\nu}\,p_{i\,\nu}$ and
 $\pt_3^{\mu}=\theta^{\mu\nu}\,p_{3\,\nu}$.
Notice that the difference between eq.~(\ref{disaster}) and 
the right hand side of eq.~(\ref{triananomaly}) is nonetheless a  
local expression; as corresponds to  
the fact that they come from  different regularizations of the same 
theory. General theorems in renormalization theory tell us that one can
retrieve eq.~(\ref{triananomaly}) from  eq.~(\ref{disaster}) by introducing
appropriate local counterterms of the field $a_{\mu}$ and its derivatives.
We shall not be concerned with the actual value of these counterterms,
but we shall point out that the coefficient of the term
\begin{equation}
\traza\Big([\T^{a_1},\T^{a_2}]\T^{a_3}\Big)\,\ee_{\muo\mutw\rho\sigma}\,p_1^{\rho}\,p_2^{\sigma}
\,\theta_{\alpha\beta}\,p_{1}^{\alpha}\,p_{2}^{\beta}
\label{theterm}
\end{equation}
is not same in both equations. Hence, there must exist a local polynomial
of $a_{\mu}$ (and its derivatives) whose BRS variation yields a contribution 
proportional to the expression in eq.~({\ref{theterm}). This casts doubts on 
the $\theta$-dependent term of eq.~(\ref{triananomaly}) as being a truly 
anomalous contribution. We shall analyse this issue below.  

Eqs.~(\ref{threepoint}),~(\ref{DRtwopoint}) and~(\ref{triananomaly}) leads
to the following candidate for  anomalous BRS equation 
\begin{equation}
\begin{array}{l}
{s\Gamma[A[a,\theta],\theta]\,=\,-\frac{i}{24\pi^2}\idx\,
\ee^{\muo\mutw\muth\mufo}\,\traza\big(\partial_{\muo}\lambda\,
a_{\mutw}\partial_{\muth}a_{\mufo}\big)}\\[9pt]
{\phantom{s\Gamma[a,\theta]}
\,+\,
\frac{1}{48\pi^2}\idx\,
\ee^{\muo\mutw\muth\mufo}\,\theta^{\alpha\beta}\,\traza\big(\partial_{\muo}\lambda\,
\partial_{\alpha}a_{\mutw}\partial_{\muth}\partial_{\beta}a_{\mufo}\big)\,+\,
o(a^3)\,+\,o(\theta^2)}\\[9pt]
{\phantom{s\Gamma[a,\theta]}\,=\,
-\frac{i}{24\pi^2}\idx\,
\ee^{\muo\mutw\muth\mufo}\,\traza\big(\partial_{\muo}\Lambda\star
A_{\mutw}\star\partial_{\muth}\A_{\mufo}\big)\,+\,
o(a^3)\,+\,o(\theta^2),}\\[9pt]
\end{array}
\label{wouldbeanomaly}
\end{equation}
where $\Lambda=\Lambda[\lambda,a,\theta]$ and $A_{\mu}=A_{\mu}[a,\theta]$ 
are defined in eq.~(\ref{SeibergWitten}).
And yet, for the right hand side of the previous equation to be a true anomaly,
we must show that there is no integrated $*$-polynomial of the 
noncommutative field $A(a,\theta)$ and its derivatives --let us call it 
$\Gamma_{ct}[A,\theta]$-- such that its BRS 
variation, $s\Gamma_{ct}[A,\theta]$, 
is, upon applying the Seiberg-Witten map, equal to the right hand side of 
eq.~(\ref{wouldbeanomaly}) up to first order in $\theta$ and up 
to two fields $a_{\mu}$. If only a  renormalizable by power-counting at the
noncommutative level $\Gamma_{ct}[A,\theta]$ is allowed, then there is only one
$\Gamma_{ct}[A,\theta]$ that might do the job, namely 
\begin{displaymath}
\Gamma_{ct}[A[a,\theta],\theta]\,=\,c\,\idx\,
\ee^{\muo\mutw\muth\mufo}\,\,\traza\big(\partial_{\muo}
A_{\mutw}\star A_{\muth}\star A_{\mufo}\big).
\end{displaymath}
$c$ is an appropriate number and, again, 
$\Lambda=\Lambda[\lambda,a,\theta]$ and $A_{\mu}=A_{\mu}[a,\theta]$ are 
as in eq.~(\ref{SeibergWitten}). Unfortunately, for this 
$\Gamma_{ct}[A[a,\theta],\theta]$, we have
\begin{displaymath}
s\Gamma_{ct}[A[a,\theta],\theta]\,=\,
s_{\star}\Gamma_{ct}[A[a,\theta],\theta]\,=\,0\,+\,
o(A^3),
\end{displaymath}
where $s$ and $s_{\star}$ are defined by eqs.~(\ref{BRStrans}) 
and~(\ref{NCBRStrans}), respectively. 
Hence, if we want to save renormalizability by power-counting at the noncommutative level, the only way that the right hand side of eq~(\ref{wouldbeanomaly}) 
would vanish is that 
\begin{equation}
\traza\,[T^a, T^b]\, T^c\,=\,0\quad\mbox{ and}\quad \traza\,\{T^a,T^b\}\,T^c\,=\,0.
\label{nccancel}
\end{equation}
And thus, unlike in ordinary space-time, the theories we are considering 
present a breach of gauge invariance even if the ordinary condition 
\begin{equation}
\traza\,\{T^a,T^b\}\,T^c\,=\,0
\label{cancel}
\end{equation}
is satisfied by the representation of the simple gauge group carried by the matter content of the theory.  This result leads immediately to the conclusion 
that the $SU(2)$ part of the noncommutative standard model of 
ref.~\cite{Calmet:2001na} and the noncommutative $SU(5)$ and $SO(10)$ models 
of ref.~\cite{Aschieri:2002mc} cannot be rendered anomaly free, if 
power-counting renormalizability is not given up at the noncommutative 
level. However, to demand that  noncommutative field theories 
--at least if they are defined by means of the Seiberg-Witten-- be renormalizable by power-counting seems to be too strong a requirement and  not in keeping
with current ideas on the renormalizability of gauge theories. Indeed, on the
one hand, even noncommutative QED fails to be renormalizable by 
power-counting, as shown in ref.~\cite{Wulkenhaar:2001sq}; and, on the other 
hand, if we adopt the effective field theory viewpoint, there is nothing 
wrong with loosing power-counting renormalizability provided BRS invariance 
is preserved~\cite{Gomis:1995jp}. If we give up the 
power-counting-renormalizabity paradigm, an interesting phenomenon takes place:
the term in eq.~(\ref{wouldbeanomaly}) linear in $\theta$ and involving two gauge fields can be canceled by adding to the classical noncommutative action of
our theory an appropriate counterterm,
$\Gamma_{ct}[A[a,\theta],\theta]$. It is not difficult to show that
\begin{equation}
\begin{array}{l}
{\Gamma_{ct}[A[a,\theta],\theta]\,=\,-\frac{1}{48\pi^2}\,\idx\,
\ee^{\muo\mutw\muth\mufo}\,\theta^{\alpha\beta}\,\traza\,\big(
\partial_{\alpha}\partial_{\muo}
A_{\mutw}\star\partial_{\muth} A_{\mufo}\star A_{\beta}\big)}\\[9pt]
{\phantom{\Gamma_{ct}[A[a,\theta],\theta]}\,=\,
-\frac{1}{48\pi^2}\,\idx\,
\ee^{\muo\mutw\muth\mufo}\,\theta^{\alpha\beta}\,\traza\,\big(
\partial_{\alpha}\partial_{\muo}
a_{\mutw}\partial_{\muth} a_{\mufo}a_{\beta}\big)\,+\,o(\theta^2).}\\[9pt]
\end{array}.
\label{begood}
\end{equation}
satisfies
\begin{equation}
s\Gamma_{ct}[A[a,\theta],\theta]\,=\,
-\frac{1}{48\pi^2}\idx\,
\ee^{\muo\mutw\muth\mufo}\,\theta^{\alpha\beta}\,\traza\big(\partial_{\muo}\lambda\,
\partial_{\alpha}a_{\mutw}\partial_{\muth}\partial_{\beta}a_{\mufo}\big)\,+\,
o(a^3)\,+\,o(\theta^2).
\end{equation}
Hence, we may define a new renormalized action
\begin{equation}
\Gamma_{new}[A[a,\theta],\theta]\,=\,\Gamma[A[a,\theta],\theta]\,+\,\Gamma_{ct}[A[a,\theta],\theta]
\end{equation}
satisfying 
\begin{equation}
s\Gamma_{new}[A[a,\theta],\theta]\,=\,-\frac{i}{24\pi^2}\idx\,
\ee^{\muo\mutw\muth\mufo}\,\traza\big(\partial_{\muo}\lambda\,
a_{\mutw}\partial_{\muth}a_{\mufo}\big)\,+\,o(a^3)\,+\,o(\theta^2).
\label{trueanomaly}
\end{equation}
The latter equation implies that the anomaly cancellation condition for 
the noncommutative three-point function of the gauge field $a_{\mu}$ is 
the ordinary one given in eq.~(\ref{cancel}). Now the $SU(2)$ part of 
the noncommutative standar model of ref.~\cite{Calmet:2001na} 
and the noncommutative $SU(5)$ and $SO(10)$ models 
of ref.~\cite{Aschieri:2002mc} carry no anomaly in the noncommutative  
three-point function of the ordinary gauge fields, since their fermion 
representations satisfy eq.~(\ref{cancel}). 
Whether the cancellation mechanism discussed above can be extended to 
the remaining Green functions and at any order in $\theta$ shall not be 
discussed here. Indeed, any feasible way of proving or disproving it shall
require the extension of the theorems on local BRS 
cohomology in ref.~\cite{Barnich:2000zw} to the case at hand. For 
further comments the reader is referred to the last two paragraphs of  
the Appendix. 

Let us close this section by rewriting 
eq.~(\ref{trueanomaly}) in terms of the noncommutative fields:
\begin{equation}
\begin{array}{l}
{s_{\star}\Gamma_{new}[A,\theta]
\,=\,-\frac{i}{24\pi^2}\idx\,
\ee^{\muo\mutw\muth\mufo}\,\traza\big(\partial_{\muo}\Lambda\star 
A_{\mutw}\partial_{\muth}\star A_{\mufo}\big)}\\[9pt]
{\phantom{s_{\star}\Gamma_{new}[A,\theta]}\,-\,
\frac{1}{48\pi^2}\traza\idx\,
\ee^{\muo\mutw\muth\mufo}\,\theta^{\alpha\beta}\,\traza\big(\partial_{\muo}\Lambda\star 
\partial_{\alpha}\star A_{\mutw}\partial_{\muth}\partial_{\beta}\star A_{\mufo}\big)\,+\,
o(a^3)\,+\,o(\theta^2),
}
\end{array}
\label{ncordinary}
\end{equation}
where $\Lambda=\Lambda[\lambda,a,\theta]$ and $A_{\mu}=A_{\mu}[a,\theta]$ 
as in eq.~(\ref{SeibergWitten}).

\section{The gauge anomaly and the ambiguity of the Seiberg-Witten map}

The issue we shall address in this section is the change, if any, of the
results presented in the previous section induced  by the freedom 
in choosing a concrete realization of the Seiberg-Witten map. Using 
the techniques of ref.~\cite{Brace:2001fj}, it is not difficult to show that
the most general 
solution~\cite{Asakawa:1999cu, Suo:2001ih, Aschieri:2002mc, Bichl:2001cq}  
to eqs.~(\ref{SWequations})  and~(\ref{Diracseibergeq}) --the Seiberg-Witten 
equations-- are the following:
\begin{equation}
\begin{array}{l}
{\Lambda^{(gen)}(a,\lambda,\theta)=\Lambda(a,\lambda,\theta)+
(2\,\kappa_{1}-i\kappa_1)\,\theta^{\alpha\beta} [a_{\alpha},\partial_{\beta}\lambda]
+o(\theta^2),}\\[9pt]
{A^{(gen)}_{\mu}(a,\theta) = A_{\mu}(a,\theta)+
\kappa_{3}\,\theta^{\alpha\beta}\,\cD_{\mu}(a)f_{\alpha\beta}
+\kappa_{4}\,\theta_{\mu}^{\phantom{\mu}\,\beta}\,\cD^{\rho}(a)f_{\rho\beta}
+(\kappa_2-\frac{i}{2}\kappa_{1})\theta^{\alpha\beta}\,\cD_{\mu}(a)[a_{\alpha},a_{\beta}]}\\[9pt]
{\phantom{A^{(gen)}_{\mu}(a,\theta) = A_{\mu}(a,\theta)+
\kappa_{3}\,\theta^{\alpha\beta}\,\cD_{\mu}(a)f_{\alpha\beta}
+\kappa_{4}\,\theta_{\mu}^{\phantom{\mu}\,\beta}\,\cD^{\rho}(a)f_{\rho\beta}
+(\kappa_2-\frac{i}{2}\kappa_{1})\theta^{\alpha\beta}\,\cD_{\mu}(a)}
+ o(\theta^2),}\\[9pt]
{\Psi^{(gen)}(a,\psi,\theta)=\Psi(a,\psi,\theta)+
t_{-}^{\mu\nu}(\eta,\gamma^{\sigma},\theta)\, D_{\mu}(a) D_{\nu}(a)P_{-}\psi 
+t_{+}^{\mu\nu}(\eta,\gamma^{\sigma},\theta)\, \partial_{\mu}\partial_{\nu} P_{+}\psi}\\[9pt]
{\phantom{\Psi^{(gen)}(a,\psi,\theta)=\Psi(a,\psi,\theta)}
+i\,\kappa_1\,\theta^{\alpha\beta}\cD_{\alpha}(a) a_{\beta}\,P_{-}\psi
+i\,
\kappa_2\,\theta^{\alpha\beta}[a_{\alpha},a_{\beta}]
\,P_{-}\psi
+o(\theta^2).}\\[9pt]
\end{array}
\label{SeibergWittengen}
\end{equation}
$\Lambda(a,\lambda,\theta)$, $A_{\mu}(a,\theta)$ and
$\Psi(a,\psi,\theta)$ are as in eqs.~(\ref{SeibergWitten}) 
and~(\ref{swmappsi}), respectively.
In the equation above,  
$t_{-}^{\mu\nu}(\eta,\gamma^{\sigma},\theta)$ and 
$t_{+}^{\mu\nu}(\eta,\gamma^{\sigma},\theta)$ are arbitrary Lorentz tensors
constructed out of the Minkowski metric, $\eta_{\mu\nu}$, the Moyal matrix
$\theta^{\mu\nu}$ and the Dirac matrices $\gamma_{\mu}$. These tensors take values on the Clifford algebra generated by $\gamma_{\mu}$, but their actual 
values will be of no relevance to our discussion. Note that we are taking 
for granted that $\Lambda^{(gen)}(a,\lambda,\theta)$, 
$A^{(gen)}_{\mu}(a,\theta)$, 
$\Psi^{(gen)}_L=P_{-}\Psi^{(gen)}(a,\psi,\theta)$ and 
$\Psi^{(gen)}_R=P_{+}\Psi^{(gen)}(a,\psi,\theta)$ transforms 
under parity as their ordinary counterparts do~\cite{Aschieri:2002mc}, this is 
why the Levi-Civita pseudotensor does not occur in 
eq.~(\ref{SeibergWittengen}). $\kappa_1$, $\kappa_2$, $\kappa_3$ and 
$\kappa_4$ are numbers. The requirement that 
$\Lambda^{(gen)}(a,\lambda,\theta)$ and $A^{(gen)}_{\mu}(a,\theta)$ be 
hermitian imposes obvious constraints on these numbers~\cite{Aschieri:2002mc}.

The regularized effective action, $\Gamma[A^{(gen)}[a,\theta],\theta]_{DR}$,  
of the noncommutative theory in dimensional regularization is given by the
diagrammatic expansion of the following  path integral
\begin{equation}
e^{\,i \Gamma[A^{(gen)}[a,\theta],\theta]_{DR}}\,=\,
\int\,d\bar{\psi}\,d{\psi}\;e^{\,
i
S^{(-)}_{nc,\,DR}(A^{(gen)}[a,\theta],\Psi^{(gen)}[a,\psi,\theta],\bar{\Psi}^{(gen)}[a,\bar{\psi},\theta])}.
\label{pathintone}
\end{equation}
Now $S^{(-)}_{nc,\,DR}(a,\psi,\bar{\psi})$ is obtained by substituting first  
eq.~(\ref{SeibergWittengen}) in eq.~(\ref{ncDR}) and then expanding the result in powers of $\theta$. As in  section 2, the Seiberg-Witten map in the 
$2\omega$-dimensional space of Dimensional Regularization  
is obtained by replacing each object in eq~(\ref{SeibergWittengen}) with its 
counterpart in the Dimensional Regularization scheme systematized in  
ref.~\cite{Breitenlohner:hr}. 

In this section we will not compute explicitly the Feynman diagrams with 
$a_{\mu}$ in the external legs that may give  anomalous contributions to 
the noncommutative three-point function of the latter gauge field. Rather, 
we shall take advantage of the fact that our Dimensional Regularization 
scheme satisfies  
the Quantum Action Principle --see~\cite{Breitenlohner:hr}-- and use the 
path integral as much as possible. Indeed, any formal manipulation of the 
path integral in eq.~(\ref{pathintone}) is mathematically sound when 
spelt out in terms of Feynman diagrams. Now, the following change of 
fermionic variables
\begin{equation}
\Upsilon\,=\,(\mbox{ I\!I}\,+\,\mbox{M}(a,\theta,\partial))\,\psi,
\label{upsilon}
\end{equation}
where the operator $\mbox{M}(a,\theta,\partial)$ is given by
\begin{equation}
\begin{array}{l}
{
\mbox{M}(a,\theta,\partial)=\big[-\frac{1}{2}\theta^{\alpha\beta}
a_{\alpha}\partial_{\beta}+i(\frac{1}{8}+\kappa_{2})
\,\theta^{\alpha\beta}[a_{\alpha},a_{\beta}]+
i\kappa_1 \theta^{\alpha\beta}\cD_{\alpha}(a)a_\beta+
t_{-}^{\mu\nu}(\theta)\, D_{\mu}(a) D_{\nu}(a)\big]{\rm P}_{-}}\\[9pt]
{
\phantom{\mbox{M}(a,\theta,\partial)\,=\,}
\,+\,t_{+}^{\mu\nu}(\theta)\, \partial_{\mu}\partial_{\nu}\mbox{P}_{+}, 
}\\[9pt] 
\end{array}
\label{opM}
\end{equation}
leaves invariant the path integral in eq.~(\ref{pathintone}). Hence, the
following equation holds up to first order in $\theta$:
\begin{equation}
\begin{array}{l}
{
\int\,d\bar{\psi}\,d{\psi}\;e^{\,
i
S^{(-)}_{nc,\,DR}(A^{(gen)}[a,\theta],\Psi^{(gen)}[a,\psi,\theta],\bar{\Psi}^{(gen)}[a,\bar{\psi},\theta])}=}\\[9pt]
{
\int\,d\bar{\Upsilon}\,d{\Upsilon}\;\;
\mbox{det}\;
\big[\mbox{ I\!I}\,+\,\mbox{M}(a,\theta,\partial)\big]
\;\;\mbox{det}\;
\big[\mbox{ I\!I}\,+\,\bar{\mbox{M}}(a,\theta,\partial)\big]\;\;
e^{\,
i
S^{(-)}_{nc,\,DR}
(A^{(gen)}[a,\theta],\Upsilon,\bar{\Upsilon})}
.}\\[9pt]
\end{array}
\label{pathintidentity}
\end{equation}
The determinants in the previous equation are defined as the sum
of the appropriate dimensionally regularized Feynman diagrams; 
$\mbox{M}(a,\theta,\partial)$ and $\bar{\mbox{M}}(a,\theta,\partial)$ 
being understood a perturbations of $\mbox{ I\!I}$. The operator 
$\mbox{ I\!I}\,+\,\bar{\mbox{M}}(a,\theta,\partial)$ yields the
change of $\bar{\psi}$, 
$\bar{\Upsilon}\,=\,(\mbox{ I\!I}+\bar{\mbox{M}}(a,\theta,\partial))\,\bar{\psi}$, induced by the change of $\psi$ in eq.~(\ref{upsilon}).

In Dimensional Regularization, we have
\begin{displaymath}
\mbox{det}\;
\big[\mbox{ I\!I}\,+\,\mbox{M}(a,\theta,\partial)\big]\,=\,1\,=\,
\;\;\mbox{det}\;
\big[\mbox{ I\!I}\,+\,\bar{\mbox{M}}(a,\theta,\partial)\big],
\end{displaymath}
for the diagrammatic expansions of these determinants always lead to 
integrals of the type
\begin{displaymath}
\idpo\;p_{\mu_1}\dots p_{\mu_{n}}.
\end{displaymath}
These integrals vanish in Dimensional regularization. We thus come to the conclusion
that in perturbation theory and at first order in $\theta$ 
$\Gamma[A^{(gen)}[a,\theta],\theta]_{DR}$ in eq.~(\ref{pathintone}) is also given
by the diagrammatic expansion of
\begin{equation}
\frac{1}{i}\,\ln\,\Big\{
\int\,d\bar{\Upsilon}\,d{\Upsilon}\;e^{\,
i
S^{(-)}_{1,\,DR}(A^{(gen)}[a,\theta],\Upsilon,\bar{\Upsilon},\theta)}\Big\},
\label{pathinttwo}
\end{equation}
with $S^{(-)}_{1,\,DR}(A^{(gen)}[a,\theta],\Upsilon,\bar{\Upsilon},\theta)$ being
given by the expansion of 
\begin{equation}
S^{(-)}_{nc,\,DR}(A^{(gen)}[a,\theta],\Upsilon,\bar{\Upsilon})\,=\,
\idxo\,
\bar{\Upsilon}\star\{\prslash\Psi-iA^{(gen)}_{\mu}[a,\theta]\bar{\gamma}^{\mu}\star\,P_{-}\Upsilon\}
\label{actionupsi}
\end{equation}
up to first order in $\theta$. The latter expansion reads 
\begin{equation}
S^{(-)}_{1,\,DR}(A^{(gen)}[a,\theta],\Upsilon,\bar{\Upsilon},\theta)=
\idxo\,
\bar{\Upsilon}\{\prslash\Upsilon-
(iA^{(gen)}_{\mu}[a,\theta]-\frac{1}{2}\theta^{\alpha\beta}
\partial_{\alpha}a_{\mu}\partial_{\beta})\,\bar{\gamma}^{\mu}\,P_{-}\Upsilon\}.
\label{actionupsiexp}
\end{equation}
It is understood here that $A^{(gen)}_{\mu}[a,\theta]$ is given by 
the right hand side of the second equation in eqs.~(\ref{SeibergWittengen}),
provided we forget about the $o(\theta^2)$ contributions.

In view of eqs.~(\ref{pathinttwo}),~(\ref{actionupsi}) and~
(\ref{actionupsiexp}) one concludes that 
$\Gamma[A^{(gen)}[a,\theta],\theta]_{DR}$ can be obtained from the 
diagrams contributing to the noncommutative $\U(\nn)$ with a left-handed 
fermion --see 
refs.~\cite{Gracia-Bondia:2000pz, Martin:2000qf}-- as follows:
\begin{itemize}
\item {\bf \it i)} Take a diagram contributing to the effective action 
of the noncommutative $\U(\nn)$ theory in question. Such a diagram, which
is always planar, has the generic form
\begin{displaymath}
\traza_{\U(\nn)}\,\idxo_{1}\cdots\idxo_{n}\;A_{\mu_{\pi(1)}}(x_{\pi(1)})\cdots
A_{\mu_{\pi(n)}}(x_{\pi(n)})\;\Gamma^{\muo\cdots\mu_{n}}(x_1,\cdots,x_n;\theta),
\end{displaymath}
with $\pi(1)\cdots \pi(n)$ being a appropriate permutation of $1 \cdots n$ and 
with 
\begin{displaymath}
\begin{array}{l}
{
\Gamma^{\muo\cdots\mu_{n}}(x_1,\cdots,x_n;\theta)=
(-1)^{n+1}\int\,\prod_{i=1}^{n}\,
 \frac{d^{2\omega}\!p_i}{(2\pi)^{2\omega}}\,
e^{\scriptscriptstyle -i\sum_{i=1}^{n} p_i x_i}\,
e^{\scriptscriptstyle -\frac{i}{2}\sum_{1\leq i<j<n}\theta^{\alpha\beta}p_{i\,\alpha}p_{j\,\beta}}
}\\[9pt]
{\phantom{\Gamma^{\muo\cdots\mu_{n}}}
(2\pi)^{2\omega}\delta(p_1+\cdots p_n)
\,\idqo
\frac{{\rm tr}[\qslash\,\bar{\gamma}^{\muo}\rP_{-}\,(\qslash-\poslash)\,
\bar{\gamma}^{\mutw}\rP_{-}\cdots (\qslash-\sum_{i=1}^{n-1}\pislash)\,
\bar{\gamma}^{\mu_{n}}\rP_{-}]}
{q^2\,(q-p_1)^2\cdots (q-\sum_{i=1}^{n-1}p_i)^2}.}\\[9pt]
\end{array}
\end{displaymath}
Then, expand at first order in $\theta$ the global Moyal phase of 
the diagram. Call the result {\it Diagram}.

\item {\bf\it ii)} Replace in {\it Diagram} the noncommutative 
$\U(\nn)$ field in the fundamental representation, $A_{\mu}$, which only 
occurs as a background field, with $A_{\mu}^{(gen)}$ defined in 
eq.~(\ref{SeibergWittengen}). And also replace $\traza_{\U(\nn)}$, the trace 
in the fundamental representation of $\U(\nn)$, 
with the trace  in the representation of our simple gauge group. 
Call the result {\it Diagram}, again.

\item {\bf \it iii)} Sum over all {\it Diagram}s obtained these way. Replace 
$A_{\mu}^{(gen)}$ with its value --given in eq.~(\ref{SeibergWittengen})-- in terms of $a_{\mu}$ and get rid of any contribution of order $\theta^2$.

\end{itemize}

Note that the dimensionally regularized action of the chiral noncommutative
$\U(\nn)$ gauge theory of refs.~\cite{Gracia-Bondia:2000pz, Martin:2000qf} 
is the action in eq.~(\ref{ncDR}), provided $A_{\mu}$ is a $\U(\nn)$ field in
the fundamental representation. Hence, the process just spelt out converts  
the action in eq.~(\ref{ncDR}) for $\U(\nn)$ into the action in 
eq.~(\ref{actionupsiexp}) for our compact simple group, $G$. So it is no  
wonder that {\it i)}--{\it iii)} yields the result we sought. 

After all these preparations it does not come as a surprise that, up 
to first order in $\theta$, the candidate for anomalous contribution 
to BRS variation of the renormalized action  
$\Gamma[A^{(gen)}[a,\theta],\theta]$, obtained from our regularized 
action $\Gamma[A^{(gen)}[a,\theta],\theta]_{DR}$ 
--see eqs.~(\ref{pathintone}) and~(\ref{pathinttwo})-- reads
\begin{equation}
\begin{array}{l}
{
s\Gamma[A^{(gen)}[a,\theta],\theta]\,=\,
-\frac{i}{24\pi^2}\idx\,
\ee^{\muo\mutw\muth\mufo}\,\traza\Big(\partial_{\muo}\lambda\,\big(
a_{\mutw}\partial_{\muth}a_{\mufo}-\frac{i}{2}a_{\mutw}
a_{\muth}a_{\mufo}\big)\Big)
}\\[9pt]
{
\phantom{s\Gamma[A^{(gen)}]}\,-\,
\frac{i}{24\pi^2}\idx\,
\ee^{\muo\mutw\muth\mufo}\,\traza\Big(\partial_{\muo}\delta\Lambda\,
\,\big(a_{\mutw}\partial_{\muth}a_{\mufo}-\frac{i}{2}a_{\mutw}
a_{\muth}a_{\mufo}\big) 
+\partial_{\muo}\lambda
\{\delta A_{\mutw},\partial_{\muth} a_{\mufo}\}\Big)
}\\[9pt]
{
\phantom{s\Gamma[A^{(gen)}]}\,-\,
\frac{1}{48\pi^2}\idx\,
\ee^{\muo\mutw\muth\mufo}\,\traza\Big(\partial_{\muo}\lambda\,\big(
[\delta A_{\mutw},a_{\muth}]a_{\mufo}+a_{\mutw}a_{\muth}\delta A_{\mufo}\big)\Big)
}\\[9pt]
{\phantom{s\Gamma[A^{(gen)}]}
\,+\,
\frac{1}{48\pi^2}\idx\,
\ee^{\muo\mutw\muth\mufo}\theta^{\alpha\beta}\,\traza\Big(\partial_{\muo}\lambda\,\big(
\partial_{\alpha}a_{\mutw}\partial_{\muth}\partial_{\beta} a_{\mufo}-
\frac{i}{2}\{\partial_{\alpha}a_{\mutw}\partial_{\beta}a_{\muth},a_{\mufo}\}\big)\Big).
}\\[9pt]
\end{array}
\label{completeanomaly}
\end{equation}
The symbols $\delta\Lambda$ and $\delta A_{\mu}$
stand for
\begin{displaymath}
\begin{array}{l}
{
\delta\Lambda=\frac{i}{4}\,\theta^{\alpha\beta}\,
\{\partial_{\alpha}\lambda,a_{\beta}\}+
(2\,\kappa_{2}-i\kappa_1)\,\theta^{\alpha\beta}  
[a_{\alpha},\partial_{\beta}\lambda]
\quad\quad\mbox{and}}\\[9pt]
{\delta A_{\mu} = -\frac{1}{4}
\theta^{\alpha\beta}
\{a_{\alpha},f_{\beta\mu}+\partial_{\beta} a_{\mu}\}+
\kappa_{3}\theta^{\alpha\beta}\cD_{\mu}f_{\alpha\beta}+
\kappa_{4}\theta_{\mu}^{\phantom{\mu}\,\beta}\cD^{\rho}f_{\rho\beta} 
+(\kappa_{2}-\frac{i}{2}\kappa_1)\theta^{\alpha\beta}\cD_{\mu}[a_{\alpha},a_{\beta}]
,}\\[9pt]
\end{array}
\end{displaymath}
respectively.
Obviously, the result in eq.~(\ref{completeanomaly}) can be retrieved from 
the consistent form of the $\U(\nn)$ gauge anomaly obtained 
in refs.~\cite{Gracia-Bondia:2000pz, Bonora:2000he}, by applying the 
process {\it i)}--{\it iii)} and replacing each ghost field in each diagram of the $\U(\nn)$ theory with $\Lambda^{(gen)}(a,\lambda,\theta)$ given in 
eq.~(\ref{SeibergWittengen}).

Eq.~(\ref{completeanomaly}) readily leads to  
\begin{equation}
\begin{array}{l}
{
s\Gamma[A^{(gen)}[a,\theta],\theta]\,=\,
-\frac{i}{24\pi^2}\idx\,
\ee^{\muo\mutw\muth\mufo}\,\traza\big(\partial_{\muo}\lambda\,
a_{\mutw}\partial_{\muth}a_{\mufo}\big)
}\\[9pt]
{
\phantom{s\Gamma[A^{(gen)}]}\,-\,
\frac{i\,\kappa_{4}}{24\pi^2}\idx\,
\ee^{\muo\mutw\muth\mufo}\,\traza\big(\partial_{\muo}\lambda
\{\theta_{\mu_2}^{\phantom{\mu_2}\,\beta}(\partial^2 a_{\beta}-\partial_{\beta}
\partial^{\rho}a_{\rho}),\partial_{\muth} a_{\mufo}\}\big)
}\\[9pt]
{\phantom{s\Gamma[A^{(gen)}]}
\,+\,
\frac{1}{48\pi^2}\idx\,
\ee^{\muo\mutw\muth\mufo}\theta^{\alpha\beta}\,\traza\big(\partial_{\muo}\lambda\,
\partial_{\alpha}a_{\mutw}\partial_{\muth}\partial_{\beta}a_{\mu_4}\big)
\,+\,o(a^3 \lambda).
}\\[9pt]
\end{array}
\label{completeexpan}
\end{equation}
Let us introduce next the nonrenormalizable noncommutative counterm
\begin{equation}
\begin{array}{l}
{
\Gamma^{(3)}_{ct}[A^{(gen)},\,\theta]=
\frac{i\,\kappa_{4}}{24\pi^2}\idx\,
\ee^{\muo\mutw\muth\mufo}\,\traza\big(A^{(gen)}_{\muo}\star
\{\theta_{\mu_2}^{\phantom{\mu_2}\,\beta}(\partial^2 A^{(gen)}_{\beta}-\partial_{\beta}
\partial^{\rho}A^{(gen)}_{\rho}),\partial_{\muth} A^{(gen)}_{\mufo}\}_{*}\big)
}\\[9pt]
{
\phantom{\Gamma^{(3)}_{ct}[A^{(gen)},\,\theta]=}
\frac{i\,\kappa_{4}}{24\pi^2}\idx\,
=\ee^{\muo\mutw\muth\mufo}\,\traza\big(a_{\muo}
\{\theta_{\mu_2}^{\phantom{\mu_2}\,\beta}(\partial^2 a_{\beta}-\partial_{\beta}
\partial^{\rho}a_{\rho}),\partial_{\muth} a_{\mufo}\}\big)+
o(a^4)+o(\theta^2),
}\\[9pt]
\end{array}
\end{equation}
with the notation $\{f,g\}_{*}=f\star g+g\star f$. The BRS variation of this
counterterm reads
\begin{displaymath}
s\Gamma^{(3)}_{ct}[A^{(gen)}[a,\theta],\,\theta]=
\frac{i\,\kappa_{4}}{24\pi^2}\idx\,
\ee^{\muo\mutw\muth\mufo}\,\traza\big(\partial_{\muo}\lambda
\{\theta_{\mu_2}^{\phantom{\mu_2}\,\beta}(\partial^2 a_{\beta}-\partial_{\beta}
\partial^{\rho}a_{\rho}),\partial_{\muth} a_{\mufo}\}\big)+o(a^3\lambda).
\end{displaymath}
As we did in the
previous section, we may define now a new renormalized effective action,
$\Gamma_{new}[A^{(gen)}[a,\theta],\,\theta]$, which satisfies  
eq.~(\ref{trueanomaly}):
\begin{equation}
\Gamma_{new}[A^{(gen)}[a,\theta]\,=\,
\Gamma[A^{(gen)}[a,\theta],\,\theta]\,+\,
\Gamma_{ct}[A^{(gen)}[a,\theta],\,\theta]\,+\,
\Gamma^{(3)}_{ct}[A^{(gen)}[a,\theta],\,\theta].
\label{newnewaction}
\end{equation}
$\Gamma_{ct}[A^{(gen)}[a,\theta],\,\theta]$ is obtained  by replacing 
$A$ with $A^{(gen)}$ in eq.~(\ref{begood}). For this new 
effective action $\Gamma_{new}[A^{(gen)}[a,\theta],\,\theta]$, the anomaly
cancellation condition up to order three in the number of $a_{\mu}$-fields
is the ordinary cancellation condition: 
\begin{equation}
\traza\,\{T^a,T^b\}\,T^c\,=\,0.
\label{cond}
\end{equation}
Hence, we conclude that at first order in $\theta$ and at least for the
three-point functions of the ordinary fields, the models formulated 
in refs.~\cite{Calmet:2001na} and~\cite{Aschieri:2002mc} are anomaly free.
It remains to be shown that the procedure introduced above can be successfully
implemented at first order in
$\theta$ for the remaining Green functions of $a_{\mu}$ and, then, show that
it also holds at any order in $\theta$.

Note that strictly speaking the counterterm $\Gamma^{(3)}_{ct}[A^{(gen)}[a,\theta],\,\theta]$ above is not needed, for the would-be anomalous term it
cancels vanishes upon imposing the ordinary anomaly cancellation condition. 
In general, it will suit our purposes to show that the sum of terms in the  
right hand side of eq.~(\ref{completeanomaly}) which are not set to zero 
by imposing eq.~(\ref{cond}) is BRS exact.

\section{Summary and Conclusions}

In this paper we have computed, using diagrammatic techniques and at first 
order in $\theta$, the consistent form of the gauge anomaly carried by 
the noncommutative three-point function of the ordinary gauge field  --call  
it $a_{\mu}$--    
of certain noncommutative chiral gauge theories defined by means of 
the Seiberg-Witten map. We have  
considered only noncommutative theories whose ordinary  matter content is 
a left-handed spinor carrying an arbitrary finite dimensional unitary 
representation of a given compact simple gauge group; the gauge group being
arbitrary as well. Our computations have led to the following conclusions:
\begin{itemize}

\item {\bf 1.-} If only noncommutative counterterms which are 
renormalizable by power-counting  are admitted --in an attempt to not to 
spoil power-counting renormalizability--, then,there is an anomalous 
noncommutative correction linear in $\theta$, besides the ordinary anomalous 
contribution, to the ordinary gauge field three-point function. To cancel both 
these anomalous contributions, the two conditions in eq.~(\ref{nccancel}) 
should be satisfied by the representation of the gauge group carried by the 
left-handed spinor of our theory. This is impossible. 
Hence, the ``safe'' representations and and ``safe'' 
groups~\cite{Georgi:1972bb} of ordinary  gauge theories are 
totally ``unsafe'' for noncommutative 
space-time. Actually, ``safe'' representations in the sense of ref.~\cite{Georgi:1972bb}
 always carry, for noncommutative space-time,  a gauge anomaly if the 
fermions of the theory all have  the 
same type of handedness. However, if there are both left-handed and 
right-handed fermions in the noncommutative  theory, then, anomaly freedom 
can be achieved provided the $\theta$ dependent piece of the right-handed and  
left-handed anomaly cancel each other, i.e.,
\begin{displaymath}
\traza [T_{L}^a,T_{L}^b]\,T_{L}^c=\traza [T_{R}^a,T_{R}^b]\,T_{R}^c.
\end{displaymath}
Here we have used an obvious notation. Note that the anomalous contributions 
furnished by a  
right-handed spinor is obtained by multiplying by $-1$ the right hand side of 
eq.~(\ref{wouldbeanomaly}) --the same for eq.~(\ref{completeanomaly})-- so that the anomalous cancellation conditions in the case at hand read
\begin{displaymath}
\traza \{T_{L}^a,T_{L}^b\}\,T_{L}^c-\traza \{T_{R}^a,T_{R}^b\}\,T_{R}^c=0,
\quad\mbox{and}\quad
\traza [T_{L}^a,T_{L}^b]\,T_{L}^c-\traza [T_{R}^a,T_{R}^b]\,T_{R}^c=0.
\end{displaymath}
Let us recall that the action of a noncommutative right-handed spinor cannot 
be expressed as the action of its charge conjugate left-handed spinor unless 
$\theta$ is replaced with $-\theta$ --see ref.~\cite{Aschieri:2002mc} for 
further details.

\item {\bf 2.-} If renormalizability by power-counting is given up and  
renormalizability in the broader sense of ref.~\cite{Gomis:1995jp} is 
called forth,  
then, nonrenormalizable noncommutative counterterms can be added to the
classical action to cancel any, linear in $\theta$, would-be anomalous contribution to the  
noncommutative three-point function of $a_{\mu}$ --see eqs.~(\ref{begood})--(\ref{ncordinary}) and 
eqs.~(\ref{completeexpan})-(\ref{newnewaction}). The anomaly cancellation 
condition for the ordinary gauge field noncommutative three-point function is 
then the ordinary  
one given in eq~(\ref{cond}). Hence,  within the framework of effective  
field theory, the noncommutative models of  
refs.~\cite{Calmet:2001na, Aschieri:2002mc} stand a fat chance of being
anomaly free models and thus becoming phenomenologically useful. It remains to 
show that the procedure described above can be still carried out successfully 
for the other Green functions of the theory, at least upon imposing 
eq.~(\ref{cond}). 
In other words, it remains to see whether the whole contribution linear in 
$\theta$ to the right hand side of eq.~(\ref{completeanomaly}) is 
(perhaps for the constraint in eq.~(\ref{cond}) ) BRS exact. The 
antifield formalism~\cite{Barnich:2000zw} may prove an invaluable tool 
in such a task, which is not 
altogether hopeless as shows the results presented here and the
fact that the noncommutative Chern-Simons action for a noncommutative field 
configuration is   
equal to the ordinary Chern-Simons of the ordinary field configuration 
obtained from the former by a non-singular Seiberg-Witten 
map~\cite{Grandi:2000av}.

Finally, the results presented in this paper along with the analysis in 
ref.~\cite{Wulkenhaar:2001sq} back up the suggestion made in  
ref.~\cite{Calmet:2001na} that noncommutative field theories defined by 
means of the Seiberg-Witten map must be formulated within the framework of 
effective field theories.

\end{itemize}

\begin{center}
\epsfig{file=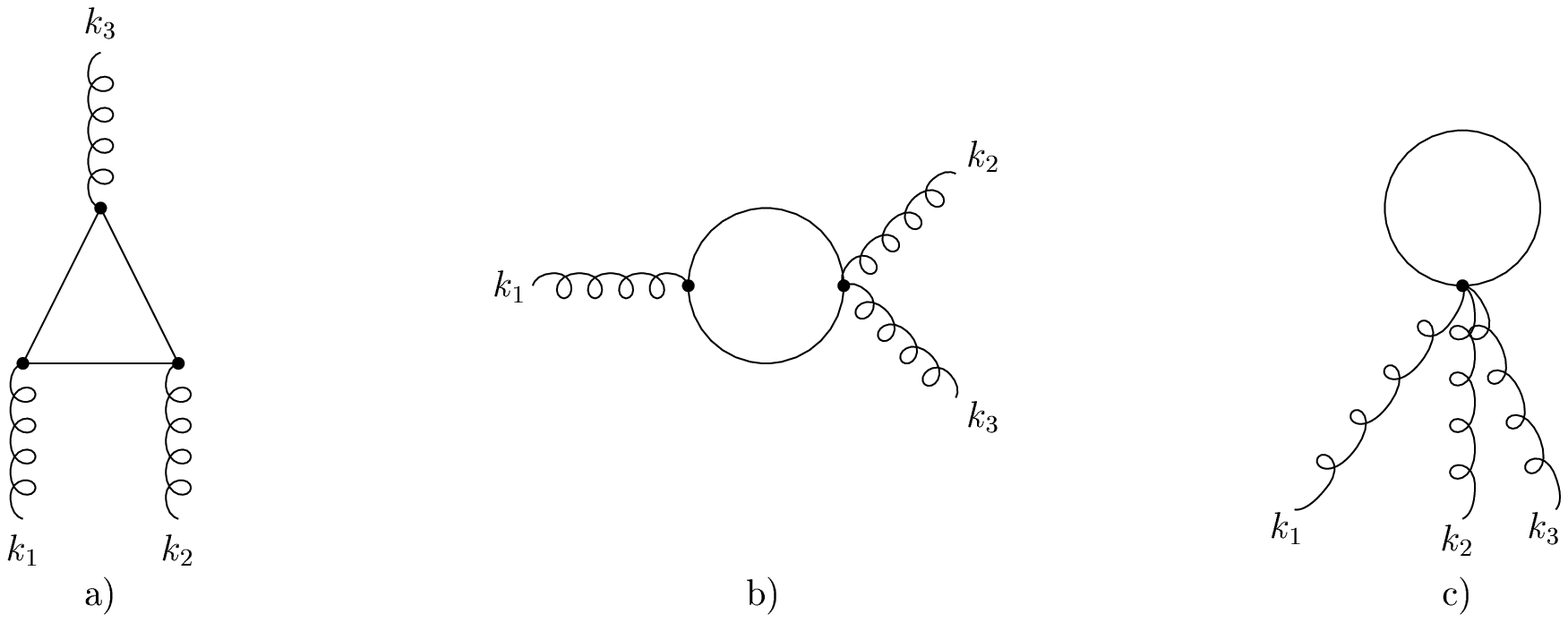,width=.60\textwidth}\\[20pt] 
{\sl Figure 1: Types of diagrams involved in computation of the noncommutative 
three-point function of $a_{\mu}$: a) triangle, b) swordfish and c) jellyfish 
diagrams.} 
\end{center}

\section{Appendix}
 
Let  
$i\pthree^{\muth}\, 
\Gamma^{(odd)\;a_1 a_2 a_3}_{DR\phantom{d}\,\;\muo\mutw\muth}
(\pone,\ptwo,\pthree)_{triangle}$ denote the contribution to the right hand
side of eq.~(\ref{DRthreepoint}) coming from the triangle-type diagrams in
fig. 1. Both $\bar{S}_{nc}$ and $\hat{S}_{nc}$ in eq.~(\ref{DRexpanded}) 
contribute to this  distribution. The bit of it which involves only vertices 
from $\bar{S}_{nc}$ is given by 
\begin{equation}
\begin{array}{l}
{\phantom{-i}\traza\Big(\{\T^{a_1},\T^{a_2}\}\T^{a_3}\Big)
\,\frac{i}{24\pi^2}\,
\ee_{\muo\mutw\rho\sigma}\,p_1^{\rho}\,p_2^{\sigma}}\\[9pt]
{
-i\,\traza\Big([\T^{a_1},\T^{a_2}]\T^{a_3}\Big)\Big\{
{\rm I}(p_1^2)\,\big(\ee_{\muo\mutw\rho\sigma}\,p_1^{\rho}\,\pt_3^{\sigma}+
\ee_{\muo\rho\sigma\tau}
\,p_1^{\rho}\,p_2^{\sigma}\,\theta_{\mutw}^{\phantom{\mutw}\,\tau}\big)
}\\[9pt]
{
\phantom{-i\,\traza\Big([\T^{a_1},\T^{a_2}]\T^{a_3}}
+{\rm I}(p_2^2)\,\big(\ee_{\muo\mutw\rho\sigma}\,p_2^{\rho}\,\pt_3^{\sigma}+
\ee_{\mutw\rho\sigma\tau}
\,p_1^{\rho}\,p_2^{\sigma}\,\theta_{\mutw}^{\phantom{\muo}\,\tau}\big)
}\\[9pt]
{\phantom{-i\,\traza\Big([\T^{a_1},\T^{a_2}]\T^{a_3}}
-\frac{i}{16\pi^2(\omega-2)}\,\big(\frac{1}{24}\big)\,
\,(\hat{p}^2_2+\hat{p}^2_3-\hat{p}_2\cdot\hat{p}_3\big)\,
\big(\ee_{\muo\mutw\rho\sigma}\,p_1^{\rho}\,\pt_3^{\sigma}+
\ee_{\muo\rho\sigma\tau}
\,p_1^{\rho}\,p_2^{\sigma}\,\theta_{\mutw}^{\phantom{\mutw}\,\tau}\big)}
\\[9pt]
{\phantom{-i\,\traza\Big([\T^{a_1},\T^{a_2}]\T^{a_3}}
-\frac{i}{16\pi^2(\omega-2)}\,\big(\frac{1}{24}\big)\,
\,(\hat{p}^2_1+\hat{p}^2_3-\hat{p}_1\cdot\hat{p}_3\big)\,
\big(\ee_{\muo\mutw\rho\sigma}\,p_2^{\rho}\,\pt_3^{\sigma}+
\ee_{\mutw\rho\sigma\tau}
\,p_1^{\rho}\,p_2^{\sigma}\,\theta_{\muo}^{\phantom{\muo}\,\tau}\big)}
\\[9pt]
{\phantom{-i\,\traza\Big([\T^{a_1},\T^{a_2}]\T^{a_3}}
+\frac{i}{16\pi^2}\,\big(\frac{1}{24}\big)\,(p^2_1+p^2_2+p_1\cdot p_2)\,
\big(
\ee_{\muo\rho\sigma\tau}
\,p_1^{\rho}\,p_2^{\sigma}\,\theta_{\mutw}^{\phantom{\mutw}\,\tau}+
\ee_{\mutw\rho\sigma\tau}
\,p_1^{\rho}\,p_2^{\sigma}\,\theta_{\muo}^{\phantom{\muo}\,\tau}
}
\\[9pt]
{\phantom{
-i\,\traza\Big([\T^{a_1},\T^{a_2}]\T^{a_3}}\phantom{
+\;\frac{i}{16\pi^2}\,\big(\frac{1}{24}\big)\,(p^2_1+p^2_2+p_1\cdot p_2)\,
\big(\ee_{\muo\rho\sigma\tau}
\,p_1^{\rho}\,p_2^{\sigma}\,\theta_{\mutw}^{\phantom{\mutw}\,\tau}}
-\ee_{\muo\mutw\rho\sigma}\,p_3^{\rho}\,\pt_3^{\sigma}\big)
}
\\[9pt]
{\phantom{-i\,\traza\Big([\T^{a_1},\T^{a_2}]\T^{a_3}}
+\frac{i}{16\pi^2}\,\big(\frac{1}{4}\big)\,
\ee_{\muo\mutw\rho\sigma}\,p_1^{\rho}\,p_2^{\sigma}
\,\theta_{\alpha\beta}\,p_{1}^{\alpha}\,p_{2}^{\beta}
\Big\}+o(\omega-2),}
\\[9pt]
\end{array}
\label{resultsone}
\end{equation}
whereas the part of it which involves only one vertex from $\hat{S}_{nc}$
reads
\begin{equation}
\begin{array}{l}
{-i\,\traza\Big([\T^{a_1},\T^{a_2}]\T^{a_3}\Big)\Big\{
\frac{i}{16\pi^2(\omega-2)}\,\big(\frac{1}{24}\big)\,
\,(\hat{p}^2_2+\hat{p}^2_3-\hat{p}_2\cdot\hat{p}_3\big)\,
\big(\ee_{\muo\mutw\rho\sigma}\,p_1^{\rho}\,\pt_3^{\sigma}+
\ee_{\muo\rho\sigma\tau}
\,p_1^{\rho}\,p_2^{\sigma}\,\theta_{\mutw}^{\phantom{\mutw}\,\tau}\big)}
\\[9pt]
{\phantom{-i\,\traza\Big([\T^{a_1},\T^{a_2}]\T^{a_3}}
+\frac{i}{16\pi^2(\omega-2)}\,\big(\frac{1}{24}\big)\,
\,(\hat{p}^2_1+\hat{p}^2_3-\hat{p}_1\cdot\hat{p}_3\big)\,
\big(\ee_{\muo\mutw\rho\sigma}\,p_2^{\rho}\,\pt_3^{\sigma}+
\ee_{\mutw\rho\sigma\tau}
\,p_1^{\rho}\,p_2^{\sigma}\,\theta_{\muo}^{\phantom{\muo}\,\tau}\big)}
\\[9pt]
{\phantom{-i\,\traza\Big([\T^{a_1},\T^{a_2}]\T^{a_3}}
-\frac{i}{16\pi^2}\,\big(\frac{1}{24}\big)\,(p^2_1+p^2_2+p_1\cdot p_2)\,
\big(
\ee_{\muo\rho\sigma\tau}
\,p_1^{\rho}\,p_2^{\sigma}\,\theta_{\mutw}^{\phantom{\mutw}\,\tau}+
\ee_{\mutw\rho\sigma\tau}
\,p_1^{\rho}\,p_2^{\sigma}\,\theta_{\muo}^{\phantom{\muo}\,\tau}
}
\\[9pt]
{\phantom{
-i\,\traza\Big([\T^{a_1},\T^{a_2}]\T^{a_3}}\phantom{
-\;\frac{i}{16\pi^2}\,\big(\frac{1}{24}\big)\,(p^2_1+p^2_2+p_1\cdot p_2)\,
\big(\ee_{\muo\rho\sigma\tau}
\,p_1^{\rho}\,p_2^{\sigma}\,\theta_{\mutw}^{\phantom{\mutw}\,\tau}}
-\ee_{\muo\mutw\rho\sigma}\,p_3^{\rho}\,\pt_3^{\sigma}\big)
}
\\[9pt]
{\phantom{-i\,\traza\Big([\T^{a_1},\T^{a_2}]\T^{a_3}}
+\frac{i}{16\pi^2}\,\big(\frac{1}{12}\big)\,
\ee_{\muo\mutw\rho\sigma}\,p_1^{\rho}\,p_2^{\sigma}
\,\theta_{\alpha\beta}\,p_{1}^{\alpha}\,p_{2}^{\beta}
\Big\}+o(\omega-2).}
\\[9pt]
\end{array}
\label{resultstwo}
\end{equation}
We are using the notation $\bar{p}_i^{\mu}=\bar{g}^{\mu\nu}\,p_{i\,\nu}$, 
$\hat{p}_i^{\mu}=\hat{g}^{\mu\nu}\,p_{i\,\nu}$,
$i=1,2,3$,  
 $\pt_3^{\mu}=\theta^{\mu\nu}\,p_{3\,\nu}$ and $p_3=-p_1-p_2$. 
${\rm I}(p^2)$ is given in eq.~(\ref{Ipedos}). 

$
i\pthree^{\muth}\,\Gamma^{(odd)\;a_1 a_2 a_3}_{DR\phantom{d}\,\;\muo\mutw\muth}
(\pone,\ptwo,\pthree)_{swordfish}$ in eq.~(\ref{DRthreepoint})
is obtained by summing over all diagrams of swordfish type 
--see b) in fig. 1. 
These diagrams only carry nonvanishing contributions
if the vertices come all from $\bar{S}_{nc}$ in eq.~(\ref{DRexpanded}). The 
sum of these swordfish-type diagrams is  
\begin{equation}
\begin{array}{l}
{
+i\,\traza\Big([\T^{a_1},\T^{a_2}]\T^{a_3}\Big)\Big\{
{\rm I}(p_1^2)\,\big(\ee_{\muo\mutw\rho\sigma}\,p_1^{\rho}\,\pt_3^{\sigma}+
\ee_{\muo\rho\sigma\tau}
\,p_1^{\rho}\,p_2^{\sigma}\,\theta_{\mutw}^{\phantom{\mutw}\,\tau}\big)
}\\[9pt]
{
\phantom{-i\,\traza\Big([\T^{a_1},\T^{a_2}]\T^{a_3}}
+{\rm I}(p_2^2)\,\big(\ee_{\muo\mutw\rho\sigma}\,p_2^{\rho}\,\pt_3^{\sigma}+
\ee_{\mutw\rho\sigma\tau}
\,p_1^{\rho}\,p_2^{\sigma}\,\theta_{\mutw}^{\phantom{\muo}\,\tau}\big)
.}\\[9pt]
\end{array}
\label{resultsthree}
\end{equation}

The jellyfish-type diagrams in c) of fig. 1, which vanish in dimensional regularization, give rise to
$
i\pthree^{\muth}\,\Gamma^{(odd)\;a_1 a_2 a_3}_{DR\phantom{d}\,\;\muo\mutw\muth}
(\pone,\ptwo,\pthree)_{swordfish}=0
$
in eq.~(\ref{results}).

The sum of the expressions in eqs.~(\ref{resultsone}),~(\ref{resultstwo}) 
and~(\ref{resultsthree}) yield the result in
eq.~(\ref{triananomaly}). This final outcome is the contribution the 
noncommutative three-point function of the ordinary field computed with the 
regularized action $S^{(-)}_{nc,\,DR}$ in eq.~(\ref{DRexpanded}). This action
is obtained from the action in eq.~(\ref{ncDR}) by using the Seiberg-Witten 
map. It is for the latter action and for the noncommutative $\U(\nn)$ groups
that the gauge anomaly has been computed without using 
the Seiberg-Witten map formalism in a number of papers  
--see refs.~\cite{Gracia-Bondia:2000pz, Bonora:2000he, Martin:2000qf}.  
Note that for the $\U(\nn)$ case 
eq.~(\ref{triananomaly}) can be obtained by applying the Seiberg-Witten map 
to the results presented in  
refs.~\cite{Gracia-Bondia:2000pz, Bonora:2000he, Martin:2000qf}. 
This is consistent
with the fact that the Seiberg-Witten map is explicitly preserved by our 
regularization procedure and  that in the aforementioned papers 
the computation of the anomaly is carried out over the space of 
$*$-polynomials of the noncommutative gauge field and its derivatives where 
$\theta^{\mu\nu}$ only occurs in the Moyal product, i.e., polynomials like 
\begin{equation}
\theta^{\alpha\beta}\,
\partial_{\alpha}\partial_{\muo}
A_{\mutw}\star\partial_{\muth} A_{\mufo}\star A_{\beta},
\label{polynomial}
\end{equation}
$A_{\mu}$ denoting the noncommutative gauge field, were not allowed: renormalizability by power-counting was a constraint.

Now, the sum of the results in eq.~(\ref{resultsone}) and 
eq.~(\ref{resultsthree}) is the ugly expression in eq.~(\ref{disaster}),
which for the $\U(\nn)$ case cannot be obtained from the results in~
\cite{Gracia-Bondia:2000pz, Bonora:2000he, Martin:2000qf} by applying the
Seiberg-Witten map technique. Note that eq.~(\ref{disaster}) corresponds
to the regularization of the theory achieved by  just using 
$\bar{S}_{nc}$ in eq.~(\ref{DRexpanded}) as the regularized action. The noncommutative ancestor of this action seems to involve  $\star$-polynomials with $\theta$ coefficients. That is to say, 
\begin{equation}
\begin{array}{l}
{\phantom{+\frac{i}{2}\theta^{\alpha\beta}}\idxo\,
\bar{\Psi}\star\{\prslash\Psi-iA_{\mu}\bar{\gamma}^{\mu}\star\,P_{-}\Psi\}}
\\[9pt]
{
+\frac{i}{2}\theta^{\alpha\beta}
\idxo\,\bar{\Psi}\star [\partial_{\alpha}A_{\beta}+ A_{\beta}\partial_{\alpha}  
-\frac{i}{2}A_{\alpha}\star A_{\beta}]\,\hat{\prslash}\,\star {\rm P}_{+}\Psi}\\[9pt]
{
-\frac{i}{2}\theta^{\alpha\beta}
\idxo\,\bar{\Psi}\star [\hat{\prslash}A_{\beta}\partial_{\alpha}
+\frac{i}{2}(\hat{\prslash}A_{\alpha}\star A_{\beta}+
                   A_{\alpha}\star \hat{\prslash}A_{\beta})+
(A_{\beta}\partial_{\alpha}+\frac{i}{2}
A_{\alpha}\star A_{\beta})\hat{\prslash}\,]\,\star{\rm P}_{-}\Psi}\\[9pt]
\end{array}
\label{cumbersome}
\end{equation}
yields $\bar{S}_{nc}$ in eq.~(\ref{DRexpanded}) upon using the Seiberg-Witten
map. One may now compute the breaking of gauge invariance in the triangle 
diagrams of the theory defined by minimal subtraction of the triangle 
diagrams of the action in eq.~(\ref{cumbersome}). This breaking is equal to
\begin{equation}
\begin{array}{c} 
{
\frac{i}{24\pi^2}\Big(\traza\, \T^{a_1}\T^{a_2}\T^{a_3}\,
e^{-\frac{i}{2}\theta^{\alpha\beta}p_{1\,\alpha}p_{2\,\beta}}\,+\,
\traza\, \T^{a_2}\T^{a_1}\T^{a_3}\,
e^{\frac{i}{2}\theta^{\alpha\beta}p_{1\,\alpha}p_{2\,\beta}}\Big)\;
\ee_{\muo\mutw\rho\sigma}\,p_1^{\rho}\,p_2^{\sigma}}\\[9pt]
-{\frac{1}{16\pi^2}\big(\frac{1}{24}\big)\Big(\,
\traza\,\T^{a_1}\T^{a_2}\T^{a_3}\,
e^{-\frac{i}{2}\theta^{\alpha\beta}p_{1\,\alpha}p_{2\,\beta}}\,-\,
\traza\, \T^{a_2}\T^{a_1}\T^{a_3}\,
e^{\frac{i}{2}\theta^{\alpha\beta}p_{1\,\alpha}p_{2\,\beta}}\Big)\Big[2\,\ee_{\muo\mutw\rho\sigma}\,p_1^{\rho}\,p_2^{\sigma}
\,\theta_{\alpha\beta}p_1^{\alpha}p_2^{\beta}}\\[9pt]
{
+(
\ee_{\muo\rho\sigma\tau}
\,p_1^{\rho}\,p_2^{\sigma}\,\theta_{\mutw}^{\phantom{\mutw}\,\tau}+
\ee_{\mutw\rho\sigma\tau}
\,p_1^{\rho}\,p_2^{\sigma}\,\theta_{\muo}^{\phantom{\muo}\,\tau}
-\ee_{\muo\mutw\rho\sigma}\,p_3^{\rho}\,\pt_3^{\sigma})
(p_1^2+p_2^2+p_1\cdot p_2)\Big],}\\[9pt]
\end{array}
\label{correction}
\end{equation}
and it agrees, as it must be, up to first order in $\theta$ with the 
expression 
in eq.~(\ref{disaster}) once the latter has been minimally subtracted. To sort 
out which terms in eq.~(\ref{correction}) are truly anomalous, if any, one 
should solve the Wess-Zumino consistency condition on the space of 
$\star$-polynomials with $\theta^{\mu\nu}$ dependent coefficients; 
a problem not studied as yet. An  
instance of the $\star$-polynomials with $\theta^{\mu\nu}$ dependent 
coefficients relevant to our problem is given in eq.~(\ref{polynomial}).

\section{Acknowledgements}

The author is grateful to J.M. Gracia-Bond\'{\i}a and R. Wulkenhaar for 
discussions. Financial support from CICyT, Spain through grant No. PB98-0842 
is acknowledged. 

\vfill\eject


\begin{thebibliography}{99}

\bibitem{Seiberg:1999vs}
N.~Seiberg and E.~Witten,
JHEP {\bf 9909}, 032 (1999)
[arXiv:hep-th/9908142].

\bibitem{Bytsko:2000di}
A.~G.~Bytsko,
JHEP {\bf 0101}, 020 (2001)
[arXiv:hep-th/0012018].



\bibitem{Madore:2000en}
J.~Madore, S.~Schraml, P.~Schupp and J.~Wess,
Eur.\ Phys.\ J.\ C {\bf 16}, 161 (2000)
[arXiv:hep-th/0001203].

\bibitem{Jurco:2000ja}
B.~Jurco, S.~Schraml, P.~Schupp and J.~Wess,
Eur.\ Phys.\ J.\ C {\bf 17}, 521 (2000)
[arXiv:hep-th/0006246].

\bibitem{Jurco:2001my}
B.~Jurco, P.~Schupp and J.~Wess,
Nucl.\ Phys.\ B {\bf 604}, 148 (2001)
[arXiv:hep-th/0102129].



\bibitem{Jurco:2001rq}
B.~Jurco, L.~Moller, S.~Schraml, P.~Schupp and J.~Wess,
Eur.\ Phys.\ J.\ C {\bf 21}, 383 (2001)
[arXiv:hep-th/0104153].

\bibitem{Calmet:2001na}
X.~Calmet, B.~Jurco, P.~Schupp, J.~Wess and M.~Wohlgenannt,
Eur.\ Phys.\ J.\ C {\bf 23}, 363 (2002)
[arXiv:hep-ph/0111115].


\bibitem{Aschieri:2002mc}
P.~Aschieri, B.~Jurco, P.~Schupp and J.~Wess,
arXiv:hep-th/0205214.

\bibitem{Bichl:2001cq}
A.~Bichl, J.~Grimstrup, H.~Grosse, L.~Popp, M.~Schweda and R.~Wulkenhaar,
JHEP {\bf 0106}, 013 (2001)
[arXiv:hep-th/0104097].


\bibitem{Wulkenhaar:2001sq}
R.~Wulkenhaar,
JHEP {\bf 0203}, 024 (2002)
[arXiv:hep-th/0112248].


\bibitem{Dobado:jx}
A.~Dobado, A.~Gomez-Nicola, J.~P.~Maroto and J.~P.~Pelaez,
{\it  N.Y., Springer-Verlag, 1997. (Texts and Monographs in Physics)}.



\bibitem{Pich:1998xt}
A.~Pich,
arXiv:hep-ph/9806303.

\bibitem{Bichl:2002wb}
A.~A.~Bichl {\it et al.},
arXiv:hep-th/0203141.



\bibitem{Putz:2002ib}
V.~Putz and R.~Wulkenhaar,
arXiv:hep-th/0205094.

\bibitem{Okuyama:1999ig}
K.~Okuyama,
JHEP {\bf 0003}, 016 (2000)
[arXiv:hep-th/9910138].

\bibitem{Garousi:1999ch}
M.~R.~Garousi,
Nucl.\ Phys.\ B {\bf 579}, 209 (2000)
[arXiv:hep-th/9909214].

\bibitem{Mehen:2000vs}
T.~Mehen and M.~B.~Wise,
JHEP {\bf 0012}, 008 (2000)
[arXiv:hep-th/0010204].

\bibitem{Fidanza:2001qm}
S.~Fidanza,
JHEP {\bf 0206}, 016 (2002)
[arXiv:hep-th/0112027].

\bibitem{Liu:2000mj}
H.~Liu,
Nucl.\ Phys.\ B {\bf 614}, 305 (2001)
[arXiv:hep-th/0011125].

\bibitem{Okawa:2001mv}
Y.~Okawa and H.~Ooguri,
Phys.\ Rev.\ D {\bf 64}, 046009 (2001)
[arXiv:hep-th/0104036].

\bibitem{Mukhi:2001vx}
S.~Mukhi and N.~V.~Suryanarayana,
JHEP {\bf 0105}, 023 (2001)
[arXiv:hep-th/0104045].

\bibitem{Liu:2001pk}
H.~Liu and J.~Michelson,
Phys.\ Lett.\ B {\bf 518}, 143 (2001)
[arXiv:hep-th/0104139].

\bibitem{Brace:2001fj}
D.~Brace, B.~L.~Cerchiai, A.~F.~Pasqua, U.~Varadarajan and B.~Zumino,
JHEP {\bf 0106}, 047 (2001)
[arXiv:hep-th/0105192].

\bibitem{Picariello:2001mu}
M.~Picariello, A.~Quadri and S.~P.~Sorella,
JHEP {\bf 0201}, 045 (2002)
[arXiv:hep-th/0110101].

\bibitem{Barnich:2002pb}
G.~Barnich, F.~Brandt and M.~Grigoriev,
JHEP {\bf 0208}, 023 (2002)
[arXiv:hep-th/0206003].

\bibitem{Asakawa:1999cu}
T.~Asakawa and I.~Kishimoto,
JHEP {\bf 9911}, 024 (1999)
[arXiv:hep-th/9909139].

\bibitem{Jurco:2001kp}
B.~Jurco, P.~Schupp and J.~Wess,
arXiv:hep-th/0106110.

\bibitem{Jackiw:2002au}
R.~Jackiw and S.~Y.~Pi,
Phys.\ Lett.\ B {\bf 534}, 181 (2002)
[arXiv:hep-th/0201251].


\bibitem{Hashimoto:2001pc}
K.~Hashimoto and H.~Ooguri,
Phys.\ Rev.\ D {\bf 64}, 106005 (2001)
[arXiv:hep-th/0105311].

\bibitem{Kraus:2001xt}
P.~Kraus and M.~Shigemori,
JHEP {\bf 0206}, 034 (2002)
[arXiv:hep-th/0110035].

\bibitem{Polychronakos:2002pm}
A.~P.~Polychronakos,
arXiv:hep-th/0206013.


\bibitem{Colladay:1998fq}
D.~Colladay and V.~A.~Kostelecky,
Phys.\ Rev.\ D {\bf 58}, 116002 (1998)
[arXiv:hep-ph/9809521].

\bibitem{Carroll:2001ws}
S.~M.~Carroll, J.~A.~Harvey, V.~A.~Kostelecky, C.~D.~Lane and T.~Okamoto,
Phys.\ Rev.\ Lett.\  {\bf 87}, 141601 (2001)
[arXiv:hep-th/0105082].


\bibitem{Carlson:2001sw}
C.~E.~Carlson, C.~D.~Carone and R.~F.~Lebed,
Phys.\ Lett.\ B {\bf 518}, 201 (2001)
[arXiv:hep-ph/0107291].

\bibitem{Banerjee:2001un}
R.~Banerjee and S.~Ghosh,
Phys.\ Lett.\ B {\bf 533}, 162 (2002)
[arXiv:hep-th/0110177].

\bibitem{Breitenlohner:hr}
P.~Breitenlohner and D.~Maison,
Commun.\ Math.\ Phys.\  {\bf 52}, 11 (1977).

\bibitem{Alvarez-Gaume:1983cs}
L.~Alvarez-Gaume and P.~Ginsparg,
Nucl.\ Phys.\ B {\bf 243}, 449 (1984).


\bibitem{Martin:1999cc}
C.~P.~Martin and D.~Sanchez-Ruiz,
Nucl.\ Phys.\ B {\bf 572}, 387 (2000)
[arXiv:hep-th/9905076].


\bibitem{Sanchez-Ruiz:2002xc}
D.~Sanchez-Ruiz,
arXiv:hep-th/0209023.

\bibitem{Gross:pv}
D.~J.~Gross and R.~Jackiw,
Phys.\ Rev.\ D {\bf 6}, 477 (1972).

\bibitem{Gomis:1995jp}
J.~Gomis and S.~Weinberg,
Nucl.\ Phys.\ B {\bf 469}, 473 (1996)
[arXiv:hep-th/9510087].

\bibitem{Barnich:2000zw}
G.~Barnich, F.~Brandt and M.~Henneaux,
Phys.\ Rept.\  {\bf 338}, 439 (2000)
[arXiv:hep-th/0002245].

\bibitem{Suo:2001ih}
B.~Suo, P.~Wang and L.~Zhao,
Commun.\ Theor.\ Phys.\  {\bf 37}, 571 (2002)
[arXiv:hep-th/0111006].

\bibitem{Gracia-Bondia:2000pz}
J.~M.~Gracia-Bondia and C.~P.~Martin,
Phys.\ Lett.\ B {\bf 479}, 321 (2000)
[arXiv:hep-th/0002171].


\bibitem{Martin:2000qf}
C.~P.~Martin,
J.\ Phys.\ A {\bf 34}, 9037 (2001)
[arXiv:hep-th/0008126].


\bibitem{Bonora:2000he}
L.~Bonora, M.~Schnabl and A.~Tomasiello,
Phys.\ Lett.\ B {\bf 485}, 311 (2000)
[arXiv:hep-th/0002210].







\bibitem{Georgi:1972bb}
H.~Georgi and S.~L.~Glashow,
Phys.\ Rev.\ D {\bf 6}, 429 (1972).


\bibitem{Grandi:2000av}
N.~Grandi and G.~A.~Silva,
Phys.\ Lett.\ B {\bf 507}, 345 (2001)
[arXiv:hep-th/0010113].




\end{thebibliography}
\end{document}